\documentclass[showpacs,amsmath,amsfonts,amssymb,aps,superscriptaddress]{revtex4}

\usepackage{makecell, tabularx}
\usepackage[dvipsnames]{xcolor}
\usepackage{comment}
\usepackage{enumitem}
\usepackage{graphicx}
\usepackage{float}
\usepackage{dcolumn}
\usepackage{bm}
\usepackage{mathrsfs}
\usepackage{amsmath}

\usepackage{amsfonts}
\usepackage{dsfont}

\begin{document}

\title{A Unified Spectral Approach for Quasinormal Modes of Lee-Wick Black Holes}
\author{Davide Batic}
\email{davide.batic@ku.ac.ae}
\affiliation{
Mathematics Department, Khalifa University of Science and Technology, PO Box 127788, Abu Dhabi, United Arab Emirates}
\author{Denys Dutykh}
\email{denys.dutykh@ku.ac.ae}
\affiliation{
Mathematics Department, Khalifa University of Science and Technology, PO Box 127788, Abu Dhabi, United Arab Emirates}
\affiliation{Causal Dynamics Pty Ltd, Perth, Australia}
\author{Breno L. Giacchini}
\email{breno.giacchini@matfyz.cuni.cz}
\affiliation{Institute of Theoretical Physics, Faculty of Mathematics and Physics, Charles University, V Hole{\v s}ovi{\v c}k{\'a}ch 2, 180 00 Prague 8, Czech Republic
}
\date{\today}

\begin{abstract}
In this paper, we undertake a comprehensive examination of quasinormal modes linked to Lee-Wick black holes, delving into scalar, electromagnetic, and gravitational perturbations using the spectral method. Such black holes can display a rich structure of horizons, and our analysis considers all the representative scenarios, including extremal and non-extremal situations. In particular, we show that purely imaginary quasinormal modes emerge for extremal and near-extremal configurations, suggesting a rapid return to equilibrium without oscillation.
\end{abstract}
\maketitle

\section{Introduction}

The perturbative approach to quantum gravity has attracted considerable attention in recent years thanks to new insights into the problem of higher derivatives and the related ghosts (see, \textit{e.g.},~\cite{Shapiro:2022ugk} for an introduction). Several models and mechanisms have been proposed to conciliate renormalizability and unitarity in quantum gravity, among which we mention the formulation of higher-derivative gravity models with complex poles in the propagator---also known as Lee-Wick gravity~\cite{Modesto:2015ozb, Modesto:2016ofr}. Simply put, the higher derivatives in the action guarantee the improved behaviour of the propagator in the UV, making the theory super-renormalizable~\cite{Asorey:1996hz}, while the ghost-like poles occurring in complex conjugate pairs yield a unitary $S$-matrix~\cite{Modesto:2015ozb}, similarly to the original proposal by Lee and Wick~\cite{Lee:1969fy, Lee:1970iw} (see also~\cite{Cutkosky:1969fq, Anselmi:2017lia, Liu:2022gun}).

At the classical level, among the most characteristic features of the complex poles are the oscillations displayed by the linearised solutions~\cite{Modesto:2016ofr, Accioly:2016qeb, Giacchini:2016xns}. Such oscillatory contribution to the gravitational force in the low-energy domain has been investigated both in theoretical and experimental research~\cite{Accioly:2016etf, Boos:2018bhd, Perivolaropoulos:2016ucs, Antoniou:2017mhs, Krishak:2020opb}. Nonetheless, higher derivatives make the study of the exact solutions of the complete field equations quite a challenging task. Therefore, a phenomenological model of Lee-Wick black holes (BHs)  incorporating the effect of the oscillations was proposed in~\cite{Bambi2017PLB}. 

These spacetimes are defined as static and spherically symmetric solutions of the effective field equations
\begin{equation}
\label{Gff}
G^{\mu} {}_{\nu} = 8 \pi G \, \tilde{T}^{\mu} {}_{\nu},
\end{equation}
where $G_{\mu\nu}$ is the Einstein tensor, $G$ is the Newton constant and
\begin{equation}
\label{effT}
\tilde{T}^{\mu} {}_\nu = \text{diag}(-\rho , p_r , p_\theta , p_\theta ) 
\end{equation}
is a model-dependent effective energy-momentum tensor which would mimic the effect of the higher derivatives~\cite{Bambi2017PLB}. In the case of a sixth-derivative Lee-Wick gravity model defined by the action
\begin{equation}
\label{action}
S = \frac{1}{16 \pi G} \int d^4 x \sqrt{-g} \left[ R + G_{\mu\nu} \left( \alpha_1 + \alpha_2 \Box \right)  R^{\mu\nu} \right], \qquad 
\alpha_2  > 0, \qquad -2 \sqrt{\alpha_2} < \alpha_1  < 2 \sqrt{\alpha_2},
\end{equation}
the corresponding effective source is given by~\cite{BreTib2,Bambi2017PLB,Giacchini2023}
\begin{equation}\label{effsource}
  \rho(r)  =  \frac{M (a^2+b^2)^2}{8 \pi ab } \frac{e^{-ar}\sin(b r) }{r},
\end{equation}
where $M$ is the total mass of the source. 
The parameters $a$ and $b$ have dimensions of the inverse length and are related to the couplings in action through
\begin{equation}\label{Rel2}
  a^2 = \frac{2 \sqrt{\alpha_2} - \alpha_1}{4 \alpha_2}, \qquad 
  b^2 = \frac{2 \sqrt{\alpha_2} + \alpha_1}{4 \alpha_2} .
\end{equation}
They represent, respectively, the real and the imaginary part of the Lee-Wick mass $m = a + i b$. Notice that the domain of the parameters $\alpha_1$ and $\alpha_2$ in~\eqref{action} ensures that $a,b\neq 0$ and $a,b\in\mathds{R}$.

The other components of~\eqref{effT} can be fixed by supplementing an effective equation of state and imposing that the continuity equation $\nabla_\mu \tilde{T}^\mu {}_\nu = 0$ is satisfied~\cite{Burzilla2023JCAP, dePaulaNetto:2023vtg, Bambi2017PLB}. This procedure is analogous to the one used for the BH inspired by noncommutative geometry~\cite{Nicolini:2005vd, Nicolini:2009gw} and other effective metrics~\cite{Giacchini2023}. For the sake of simplicity, in this work, we consider the equation of state $p_r = - \rho$.

Various aspects of Lee-Wick BHs associated with the smeared source~\eqref{effsource} have been studied, such as the curvature regularity and BH thermodynamics~\cite{Bambi2017PLB, Burzilla2023JCAP}, the structure of horizons and regimes for BH sizes~\cite{Burzilla2023JCAP}, the gravitational light deflection~\cite{Zhao:2017jmv, Buoninfante:2020qud, Zhu:2020wtp}, precession of orbits~\cite{Lin:2022wda} and a rotating generalization based on the Newman-Janis algorithm~\cite{Singh:2022tlo}. In particular, in~\cite{Burzilla2023JCAP}, it was shown that the oscillation pattern of the solution depends on the ratio $q = b/a$, which also affects the possible number of horizons and defines admissible intervals for the position of the event horizon. The actual number of horizons is determined by $q$ and the mass $M$ of the effective source. 

In the present work, we investigate how this rich structure of horizons and regimes of Lee-Wick BHs affect the quasinormal modes (QNMs)  for massless scalar and electromagnetic and gravitational perturbations. In what concerns the latter, in our study, we assume that the gravitational perturbations follow second-order differential equations, as is the case in general relativity. This approach is not incorrect, for the background geometry is obtained as a solution of effective field equations~\eqref{Gff} that have the same form as Einstein equations. However, these tensor perturbations cannot be used to investigate the stability of the solutions to the Lee-Wick gravity, which follows sixth-derivative equations of motion. The stability of solutions in models with complex modes is an important open problem, and this study can be regarded as a first step towards this most interesting case.

We close this introductory section by recalling that the Lee-Wick BHs proposed in~\cite{Bambi2017PLB, Burzilla2023JCAP} are not exact solutions of the field equations of sixth-derivative gravity, but they might reproduce some of their relevant features, e.g., curvature regularity and the existence of multiple and extreme horizons. Exact solutions to the field equations of the extended Einstein-Hilbert action, including all possible higher-derivative terms with up to six metric derivatives, have been studied in detail only recently, and it was shown that all the possible vacuum solutions admitting a Frobenius expansion around $r=0$ have a bounded Kretschmann scalar $R_{\mu\nu\alpha\beta}^2$~\cite{Giacchini:2024exc}. Although the particular model~\eqref{action} also admits singular solutions, its solution equivalent to Schwarzschild for general relativity seems to be a regular metric~\cite{Giacchini:2024exc}. The results available for the moment only concern either local aspects of the solutions or the regime of weak field~\cite{Holdom:2002xy, Giacchini:2024exc, Pawlowski:2023dda}, not being conclusive to verify if the structure of horizons described in~\cite{Burzilla2023JCAP} matches the one of the exact solutions. Nevertheless, simple BH horizons and extreme horizons do occur in exact solutions of sixth-derivative gravity~\cite{Giacchini:2024exc}, which is in line with the argument that a regular and asymptotically flat geometry must have an even number of horizons~\cite{Holdom:2002xy}.

The structure of our paper is as follows: In Sec.~\ref{Sec2} we obtain the equations of motion for the scalar, electromagnetic, and gravitational perturbations immersed in a Lee-Wick BH background in a form suitable for the application of the spectral method to calculate the QNMs. The consideration is divided into two parts, one for the non-extremal situation and another for the case of an extreme BH. The numerical methods and results are presented in Section~\ref{Sec3}. Our analysis focuses on the case $q=2$, which is a representative case of Lee-Wick BHs that displays various possibilities for a number of horizons and extremal configurations, but we also considered some cases with $q \leq 1$ that only have one BH regime. In particular, we show that purely imaginary QNMs emerge for extremal and near-extremal configurations, suggesting a rapid return to equilibrium without oscillation. Finally, in Sec.~\ref{Sec4} we draw our conclusions.

\section{Equations of motion}\label{Sec2}

\begin{figure}[ht!] 
    \includegraphics[width=0.3\textwidth]{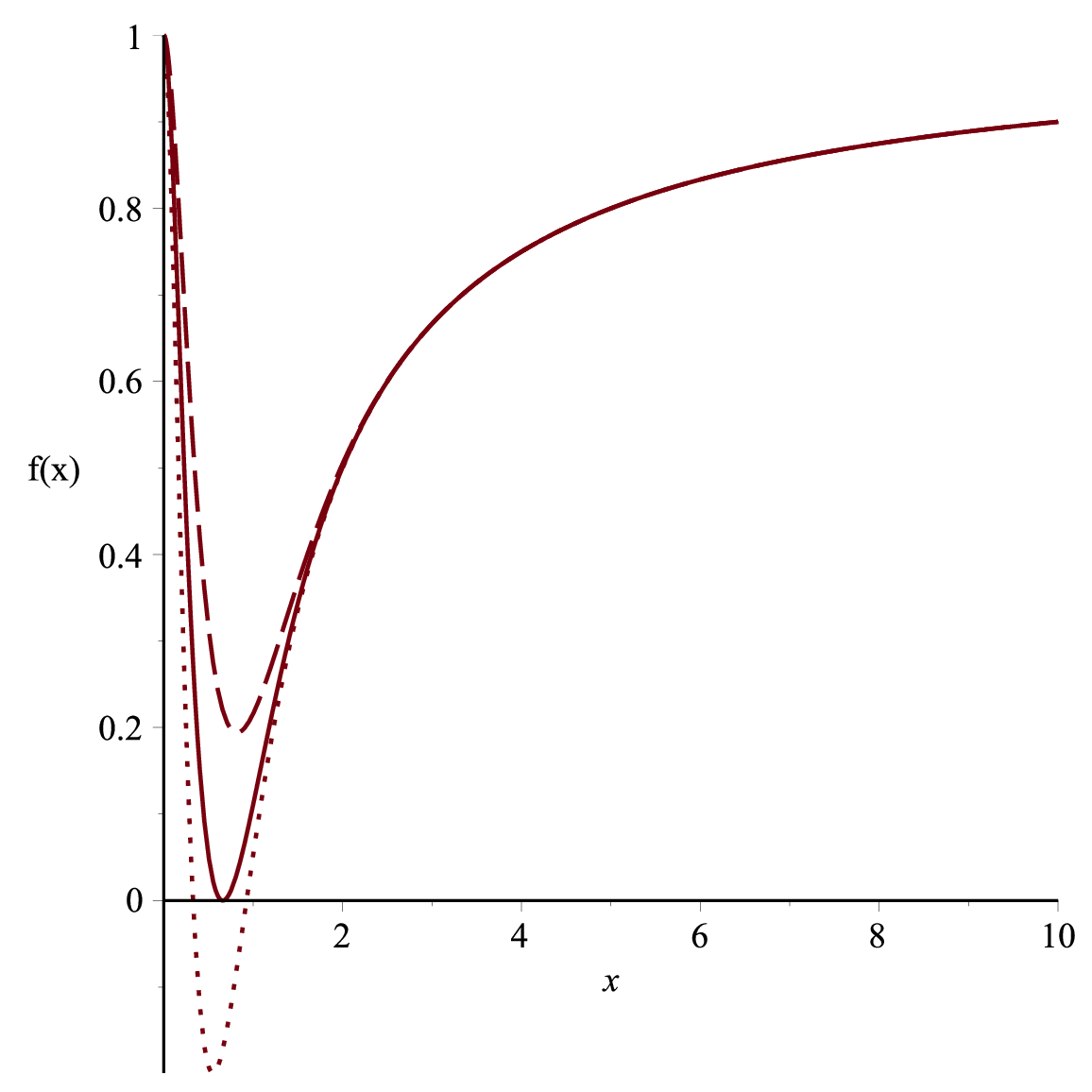}
    \includegraphics[width=0.3\textwidth]{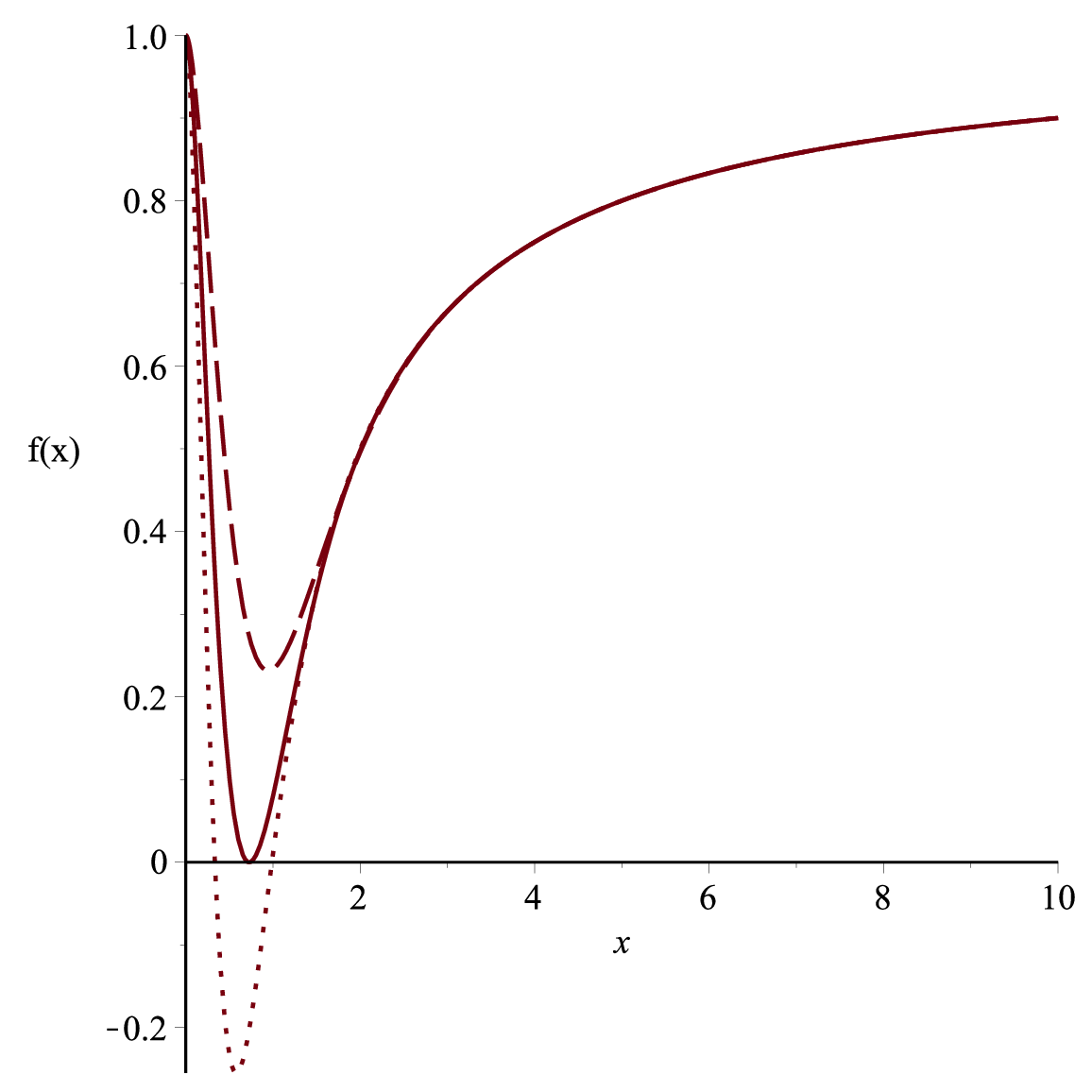}
    \includegraphics[width=0.3\textwidth]{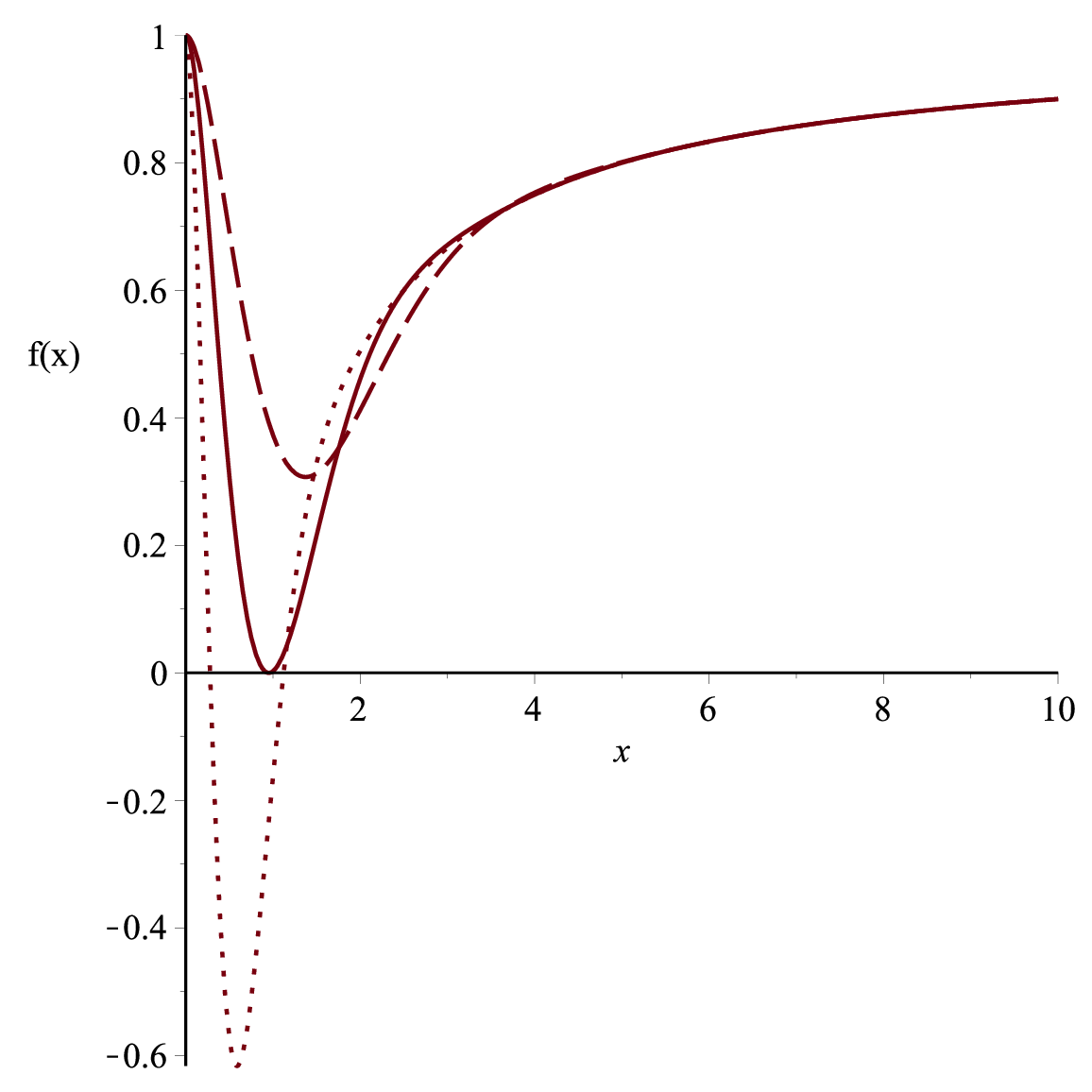}
    \includegraphics[width=0.3\textwidth]{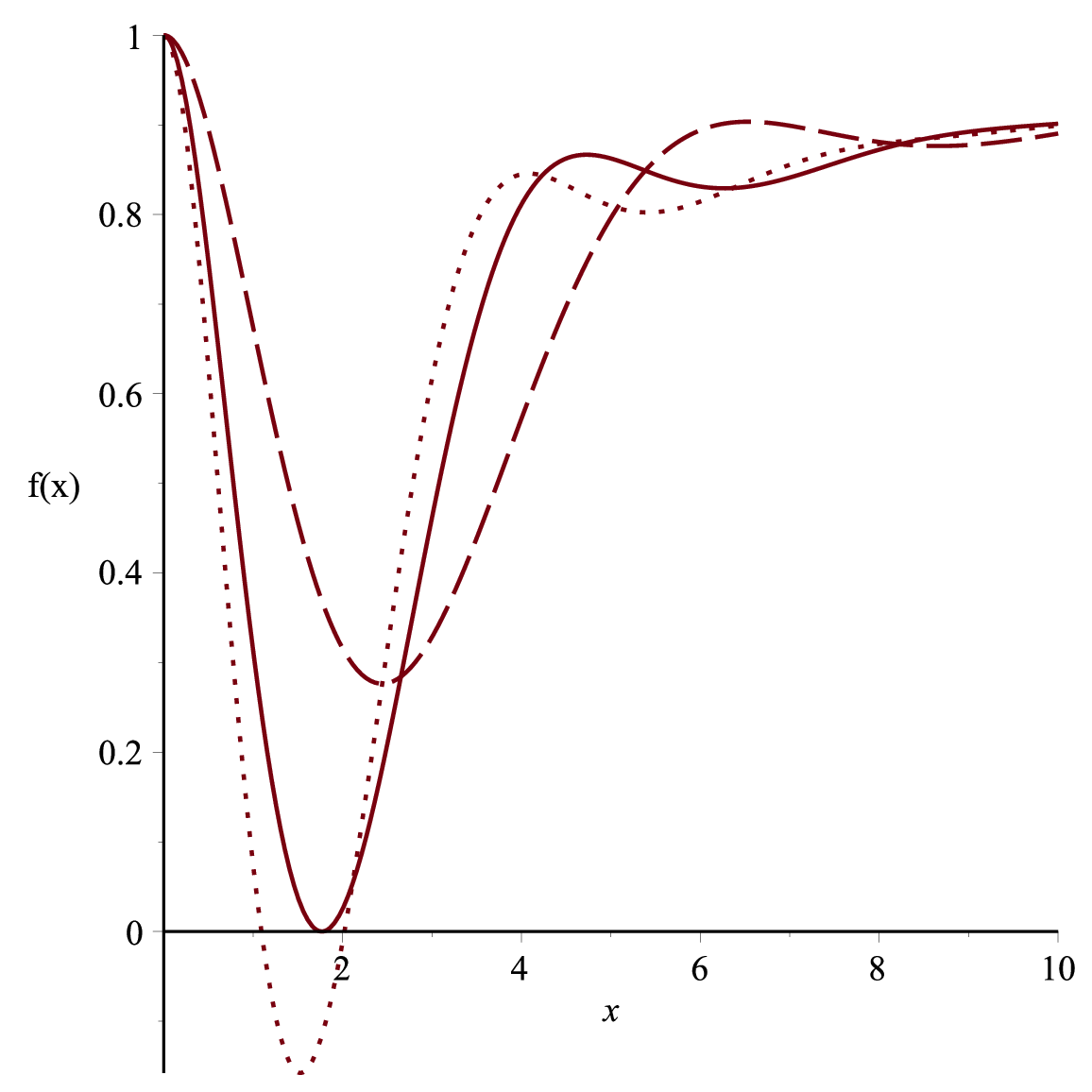}
    \includegraphics[width=0.3\textwidth]{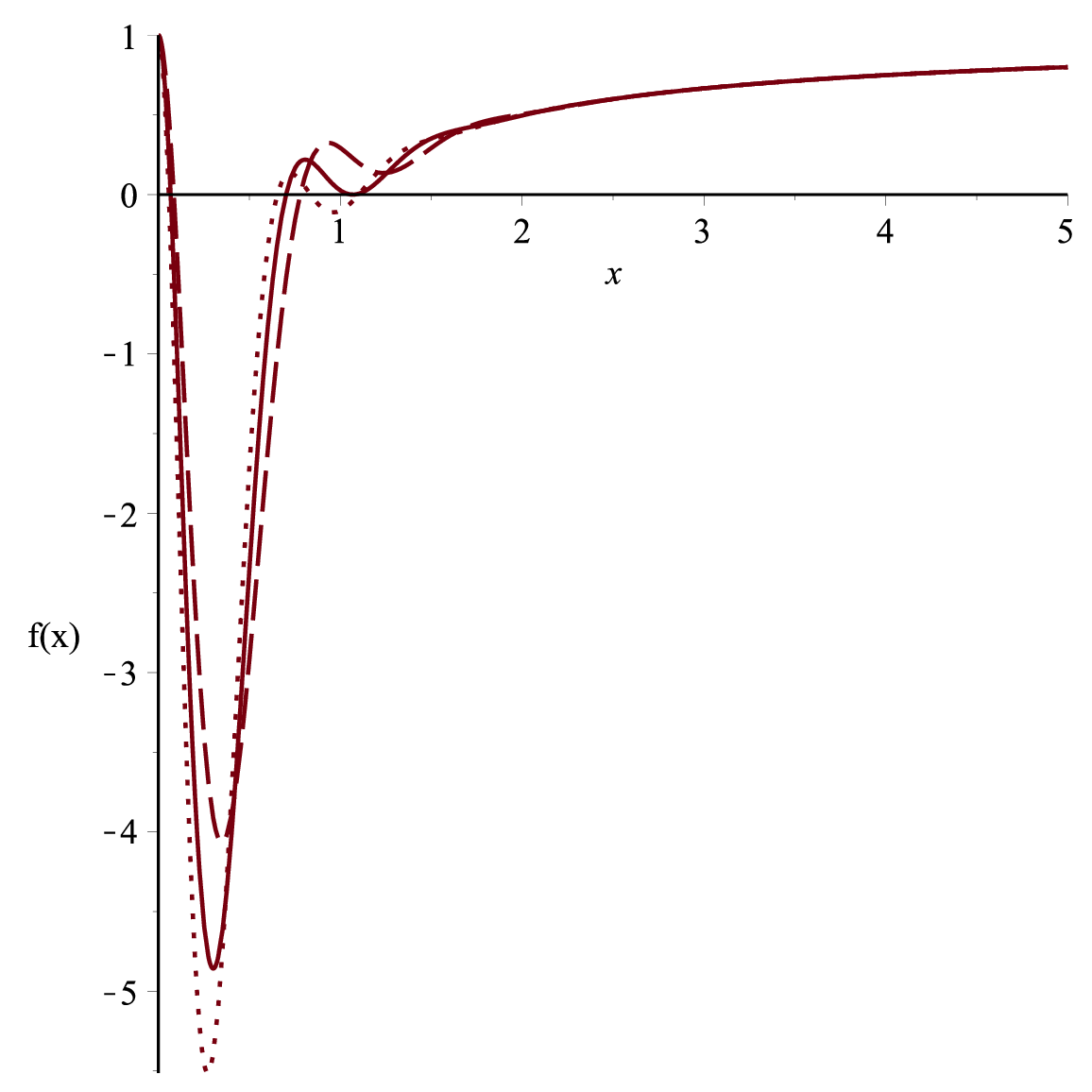}
    \includegraphics[width=0.3\textwidth]{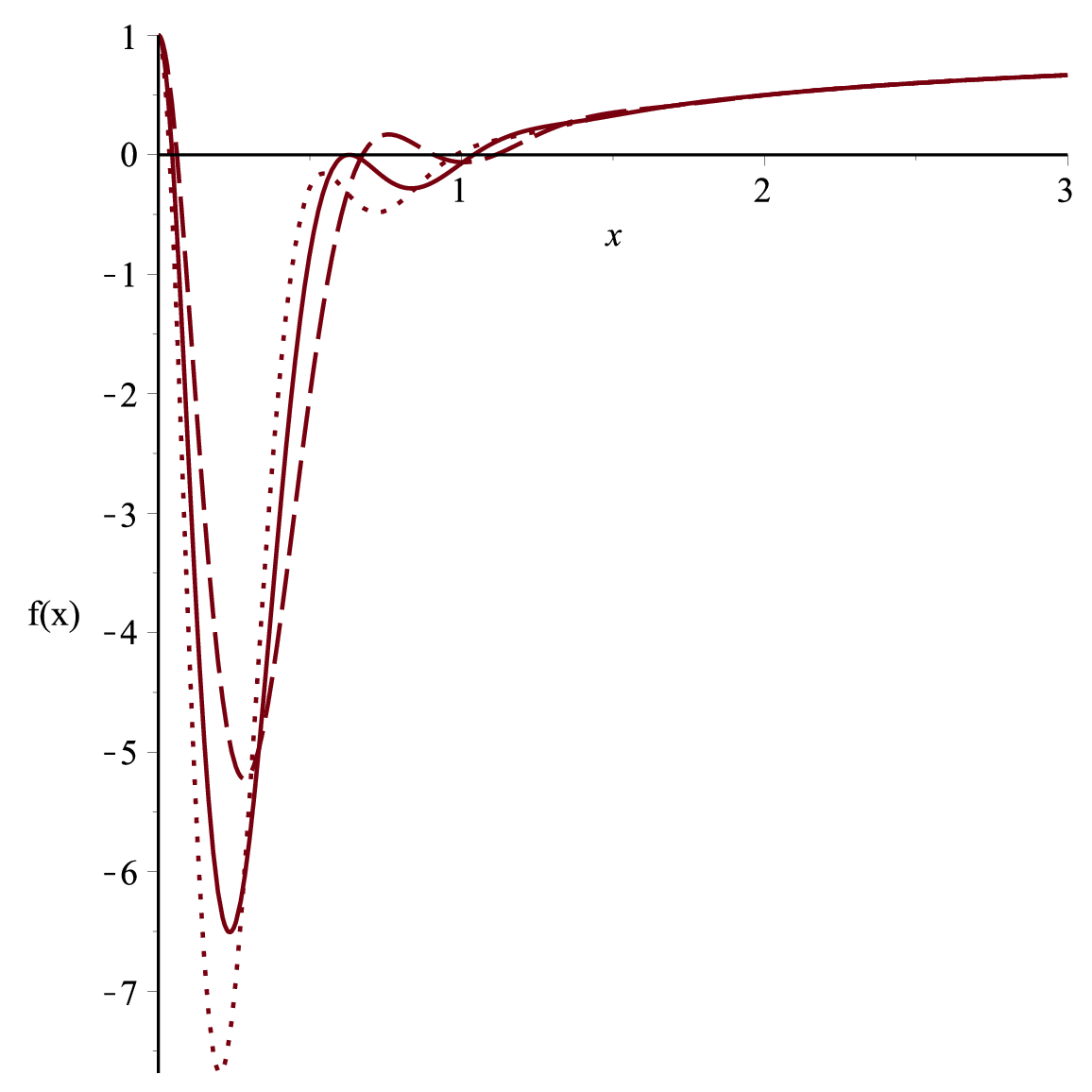}
    \caption{\label{figure01}
Plots of $f(x)$ as given by (\ref{fxfinal}) for different values of $q = \beta/\alpha$. {\bf{First row}}. Left panel: $q = 0.1$, $\alpha = 6.1$ (dotted line, two distinct horizons), $\alpha = 4.1$ (dashed line, no horizons), $\alpha = 5.08671$ (solid line, extreme case with two coinciding horizons at $x_e = 0.65978$). Central panel: $q = 0.5$, $\alpha = 4.9$ (dotted line, two distinct horizons), $\alpha = 3.0$ (dashed line, no horizons), $\alpha = 3.90504$ (solid line, extreme case with two coinciding horizons at $x_e = 0.72769$). Right panel: $q = 1$, $\alpha = 3.5$ (dotted line, two distinct horizons), $\alpha = 1.5$ (dashed line, no horizons), $\alpha = 2.16496$ (solid line, extreme case with two coinciding horizons at $x_e = 0.95306$). {\bf{Second row}}. Left panel: $q = 2$, $\alpha = 0.8$ (dotted line, two distinct horizons), $\alpha = 0.5$ (dashed line, no horizons), $\alpha = 0.69098$ (solid line, extreme case with two coinciding horizons at $x_e = 1.76785$). Central panel: $q = 2$, $\alpha = 4.5$ (dotted line, four distinct horizons), $\alpha = 3.5$ (dashed line, two distinct horizons), $\alpha = 4.04733$ (solid line, extreme case with two inner distinct horizons and two coinciding horizons at $x_e = 1.06972$). Right panel: $q = 2$, $\alpha = 6.0$ (dotted line, two distinct horizons), $\alpha = 4.3$ (dashed line, four distinct horizons), $\alpha = 5.18635$ (solid line, two coinciding inner horizons at $x_e = 0.62987$ and two distinct horizons. The event horizon is located at $x_h = 1.04013$).}
\end{figure}

We consider massless fields immersed in the Lee-Wick background associated with the effective source~\eqref{effsource}, whose line element in units where $c = G = 1$ reads  \cite{Bambi2017PLB, Burzilla2023JCAP}
\begin{equation}\label{metric}
  ds^2 = -f(r)dt^2+\frac{dr^2}{f(r)} + r^2d\vartheta^2 + r^2\sin^2{\vartheta}d\varphi^2, \quad \vartheta\in[0,\pi], \quad \varphi\in[0,2\pi),
\end{equation}
where
\begin{eqnarray}
    f(r)&=&1-\frac{2M}{r}h(r),\label{fr}\\
    h(r)&=&1-\frac{e^{-ar}}{2ab}\left\{
    b\left[2a+(a^2+b^2)r\right]\cos{(br)}
    +\left[a^2-b^2+a(a^2+b^2)r\right]\sin{(br)}\right\}.\label{hr}
\end{eqnarray}
The parameters $a$ and $b$, defined in~\eqref{Rel2}, are related to the so-called Lee-Wick ``mass'' by $m = a + ib$ with $a,b>0$. For this choice of the sign for the parameters $a$ and $b$, we refer to \cite{Burzilla2023JCAP}. Horizons exist at those values of $r$ where $f(r)$ not only vanishes but also changes signs when crossing these points. Moreover, the case of a Schwarzschild BH is recovered in the asymptotic regime $r \gg 1/a$. If we introduce the following rescaling
\begin{equation}
    x=\frac{r}{2M},\quad
    \alpha=2Ma,\quad
    \beta=2Mb,
\end{equation}
we can cast (\ref{hr}) in the form
\begin{equation}\label{fx}
    h(x)=1-e^{-\alpha x}\left[
    \left(1+\frac{\alpha^2+\beta^2}{2\alpha}x\right)\cos{(\beta x)+\frac{1}{2}\left(\frac{\alpha^2-\beta^2}{\alpha\beta}+\frac{\alpha^2+\beta^2}{\beta}x\right)\sin{(\beta x)}}.
    \right]
\end{equation}
Finally, if we introduce the ratio $q = \beta/\alpha$ as in \cite{Burzilla2023JCAP}, (\ref{fx}) becomes
\begin{equation}\label{hxfinal}
    h(x)=1-e^{-\alpha x}\left\{
    \left[1+\frac{\alpha}{2}(1+q^2)x\right]\cos{(\alpha qx)+
    \frac{1}{2q}\left[1-q^2+\alpha(1+q^2)x\right]\sin{(\alpha q x)}}
    \right\}
\end{equation}
and
\begin{equation}\label{fxfinal}
    f(x)=1-\frac{h(x)}{x}.
\end{equation}
It is rewarding to observe that for $q = 1 = \alpha$, our results for (\ref{hxfinal}) accurately replicate the function $f(x)$ as described by equation $(15)$ in \cite{Bambi2017PLB}. Figure~\ref{figure01} illustrates the complex horizon structure within the metric (\ref{metric}), highlighting how increasing $q$ leads to a greater number of zeroes in $f(x)$, thereby introducing the potential for various extreme BH configurations, as detailed in Table~\ref{tableEins}. Specifically, for $0 < q \leq 1$ and $\alpha > \alpha_e$ (with $\alpha_e$ being the extreme BH threshold), a maximum of two distinct horizons can exist, the larger of which closely resembles the Schwarzschild scenario. This observation will be used later on to validate the Spectral Method in deriving QNMs for massless scalar, electromagnetic, and gravitational perturbations. The dynamics shift markedly for $q > 1$, where distinct $\alpha$ values can lead to extreme BHs, indicating a potential significant deviation of the numerical values of the QNMs from the Schwarzschild benchmark. Additionally, Figure~\ref{figure01} (see the last panel in the second row) illustrates that horizons may merge within the outer event horizon, a phenomenon distinctly different from the traditional merging of the Cauchy and event horizons seen in extreme BHs (refer to the first panel in the second row). Therefore, we classify this unique scenario as an {\it{extreme BH of type A}}. In this case, the surface gravity $\kappa$ does not vanish at the event horizon. Conversely, we define a {\it{BH of type B}} as one where horizon coalescence occurs at the outer event horizon itself.

\begin{table}
\caption{Analysis of the real roots of the function $f(x)$ given by (\ref{fxfinal}) for several values of the parameters $q$ and $\alpha$. $\alpha_e$ denotes the corresponding value of the parameter $\alpha$ for an extreme Lee-Wick BH.} 
\begin{center}
\begin{tabular}{ | c | c | c |}
\hline
$q=\beta/\alpha$  & $\alpha$  & \mbox{Number of real roots}\\ \hline
$0.1$             & $0<\alpha<\alpha_e$      & $0$          \\ \hline
                  & $\alpha_e=5.08671$        & $1$ \mbox{double root}\\ \hline
                  & $\alpha>\alpha_e$        & $2$  \mbox{distinct roots} \\ \hline
$0.5$             & $0<\alpha<\alpha_e$      & $0$          \\ \hline
                  & $\alpha_e=3.90504$        & $1$ \mbox{double root}\\ \hline
                  & $\alpha>\alpha_e$        & $2$  \mbox{distinct roots} \\ \hline
$1.0$             & $0<\alpha<\alpha_e$      & $0$          \\ \hline
                  & $\alpha_e=2.16496$        & $1$ \mbox{double root}\\ \hline
                  & $\alpha>\alpha_e$        & $2$  \mbox{distinct roots} \\ \hline
$2.0$             & $0<\alpha<\alpha_{e,1}$      & $0$          \\ \hline
                  & $\alpha_{e,1}=0.69098$        & $1$ \mbox{double root, type B BH}\\ \hline
                  & $\alpha_{e,1}<\alpha<\alpha_{e,2}$  & $2$  \mbox{distinct roots} \\ \hline
                  & $\alpha_{e,2}=4.04733$  &  $2$  \mbox{distinct roots and } $1$ \mbox{double root, type B BH} \\ \hline
                  & $\alpha_{e,2}<\alpha<\alpha_{e,3}$ & $4$ \mbox{distinct roots}\\ \hline
                  & $\alpha_{e,3}=5.18633$ & $2$  \mbox{distinct roots and } $1$ \mbox{double root, type A BH} \\ \hline
                  & $\alpha>\alpha_{e,3}$ & $2$  \mbox{distinct roots} \\ \hline
\end{tabular}
\label{tableEins}
\end{center}
\end{table} 
In the present work, we focus our analysis on the QNMs for the nonextreme and extreme cases. Considering this, it is advantageous to implement the rescaling $z = x/x_h$, as this approach effectively maps the event horizon to $1$ in both the extreme and nonextreme cases. Hence, the coefficient function, as expressed by (\ref{fxfinal}), becomes
\begin{equation}\label{fzfinal}
    f(z)=1-\frac{h(z)}{x_h z},\quad
    h(z)=1-e^{-\alpha x_h z}\left\{
    \left[1+\frac{\alpha x_h}{2}(1+q^2)z\right]\cos{(\alpha x_h qz)+
    \frac{1}{2q}\left[1-q^2+\alpha x_h(1+q^2)z\right]\sin{(\alpha x_h z)}}
    \right\}.
\end{equation}
Given that the line element (\ref{metric}) closely resembles that of a BH inspired by noncommutative geometry~\cite{Nicolini:2005vd}, the equation for a massless Klein–Gordon field—with a time dependence expressed as $e^{-i\omega t}$ and an angular component represented by spherical harmonics—takes the form as follows \cite{Batic2019EPJC}
\begin{equation}\label{ODE01}
    f(r)\frac{d}{dr}\left(f(r)\frac{d\psi_{\omega\ell\epsilon}}{dr}\right)+\left[\omega^2-U_\epsilon(r)\right]\psi_{\omega\ell\epsilon}(r)=0,\qquad
    U_\epsilon(r)=f(r)\left[\frac{\epsilon}{r}\frac{df}{dr}+\frac{\ell(\ell+1)}{r^2}\right],\qquad
    \epsilon=1-s^2
\end{equation}
with $\ell = 0, 1, 2, \ldots$ and $\epsilon = 1$ (massless scalar perturbation $s=0$), $\epsilon = 0$ (electromagnetic perturbation $s=1$), and $\epsilon = -3$ (vector-type gravitational perturbation $s=2$). By means of the substitution $r=2Mx_h z$, the above equation can be recast in the equivalent form
\begin{equation}\label{ourODE}
    \frac{f(z)}{4}\frac{d}{dz}\left(f(z)\frac{d\psi_{\Omega\ell\epsilon}}{dz}\right)+\left[x_h^2\Omega^2-V_\epsilon(z)\right]\psi_{\Omega\ell\epsilon}(z)=0,\qquad
    V_\epsilon(z)=\frac{f(z)}{4}\left[\frac{\epsilon}{z}\frac{df}{dz}+\frac{\ell(\ell+1)}{z^2}\right],\qquad
    \Omega=M\omega
\end{equation} 
with $f(z)$ as given in (\ref{fzfinal}). Our analysis concentrates on calculating the QNMs from the specified spectral problem in (\ref{ourODE}). We denote $\Omega$ as $\Omega = \Omega_R + i\Omega_I$, with $\Omega_I < 0$ indicating time-damped perturbations. The boundary conditions ensure radial fields reflect inward radiation at the event horizon and outward at infinity, requiring us to closely inspect the asymptotic behaviour of the solution in (\ref{ourODE}) near the event horizon ($z \to 1^{+}$) and at far distances ($z \to +\infty$). For spectral method application, we adapt (\ref{ourODE}) and its boundary conditions to a $[-1, 1]$ interval, using Chebyshev polynomials to express the eigenfunctions' regular components.

\subsection{The non-extreme case}
\label{Sec2a}

This scenario examines choices of the parameter $q$ that ensure the existence of an interval for $\alpha$ where the Cauchy and event horizons remain distinct. It encompasses extreme BHs of type A, where horizon merging occurs within the event horizon, the latter not undergoing any coalescence process. For instances of $q$ and $\alpha$ values that lead to the outlined scenario, see Figure~\ref{figure01} and Table~\ref{tableEins}. In the present case, $f(z)$ exhibits a simple zero at $z = 1$. To establish the QNM boundary conditions at the event horizon and at infinity, we first need to analyze the asymptotic behaviour of the radial solution $\psi_{\Omega\ell\epsilon}$ as $z \to 1^{+}$ and as $z \to +\infty$. We can then extract the QNM boundary conditions from this asymptotic data. We split our analysis by examining the behaviour of the radial field in two different regions. 
\begin{enumerate}
\item 
{\underline{Asymptotic behaviour as $z\to 1^+$}}: Since $z=1$ is a simple zero of $f(z)$, we can construct the representation $f(z) = (z-1) g(z)$ where $g$ is an analytic function at $z=1$ with the property that $g(1) = f'(1) \neq 0$. Here, the prime symbol stands for differentiation with respect to $z$. This representation enables us to reformulate (\ref{ourODE}) in the form
    \begin{eqnarray}
        &&\frac{d^2\psi_{\Omega\ell\epsilon}}{dz^2}+p(z)\frac{d\psi_{\Omega\ell\epsilon}}{dz}+q(z)\psi_{\Omega\ell\epsilon}(z)=0,\label{ODEZ}\\
        &&p(z)=\frac{f^{'}(z)}{(z-1)g(z)},\quad
        q(z)=\frac{4x_h^2\Omega^2}{(z-1)^2 g^2(z)}-\frac{1}{(z-1)g(z)}\left[\frac{\epsilon}{z}f^{'}(z)+\frac{\ell(\ell+1)}{z^2}\right].
    \end{eqnarray}
Since $p$ and $q$ have poles of order one and two at $z = 1$, respectively, this point is a regular singular point of (\ref{ODEZ}), according to Frobenius theory \cite{Ince1956}. Hence, we can construct solutions of the form
\begin{equation}
        \psi_{\Omega\ell\epsilon}(z) = (z-1)^\rho\sum_{\kappa=0}^\infty a_\kappa(z-1)^\kappa.
\end{equation}
The leading behavior at $z = 1$ is represented by the term $(z-1)^\rho$ where $\rho$ is determined by the indicial equation
\begin{equation}\label{indicial}
        \rho(\rho-1) + P_0\rho + Q_0 = 0
\end{equation}
with
\begin{equation}
        P_0 = \lim_{z \to 1}(z-1)p(z) = 1, \qquad
        Q_0=\lim_{z \to 1}(z-1)^2 q(z) = \left(\frac{2x_h\Omega}{f^{'}(1)}\right)^2.
\end{equation}
The roots of (\ref{indicial}) are $\rho_\pm = \pm 2i x_h\Omega/f^{'}(1)$ and the correct QNM boundary condition at $z = 1$ reads
\begin{equation}\label{QNMBCz1}
\psi_{\Omega\ell\epsilon}\underset{{z\to 1^+}}{\longrightarrow} (z-1)^{-2i x_h\widetilde{\alpha}\Omega},\quad\widetilde{\alpha}=\frac{1}{f^{'}(1)}.
\end{equation}
In order to evaluate $\widetilde{\alpha}$, we start by observing that if we solve the equation $f(1) = 0$ with respect to the exponential function, we obtain 
\begin{equation}\label{exp}
e^{-\alpha x_h}=\frac{2q(1-x_h)}{A_0(\alpha,q)\cos{(\alpha x_h q)}+A_1(\alpha,q)\sin{(\alpha x_h q)}}
\end{equation}
with
\begin{equation}
A_0(\alpha,q)=\alpha q x_h(1+q^2)+2q,\quad
A_1(\alpha,q)=\alpha x_h(1+q^2)+1-q^2.
\end{equation}
Computing $f^{'}(1)$ and making use of (\ref{exp}) yields 
\begin{equation}\label{alphatilde}
\widetilde{\alpha}=\frac{A_0(\alpha,q)\cos{(\alpha x_h q)}+A_1(\alpha,q)\sin{(\alpha x_h q)}}{A_0(\alpha,q)\cos{(\alpha x_h q)}+B_0(\alpha,q)\sin{(\alpha x_h q)}},
\end{equation}
where
\begin{equation}
B_0(\alpha,q)=\alpha x_h(1+q^2)\left[\alpha(1+q^2)(x_h-1)+1\right]+1-q^2.    
\end{equation}

\item 
{\underline{Asymptotic behaviour as $z\to+\infty$}}:  We first rewrite (\ref{ourODE}) as
\begin{equation}\label{ODEas}
        \frac{d^2\psi_{\Omega\ell\epsilon}}{dz^2} + P(z)\frac{d\psi_{\Omega\ell\epsilon}}{dz} + Q(z)\psi_{\Omega\ell\epsilon}(z) = 0, \qquad
        P(z) = \frac{f^{'}(z)}{f(z)}, \qquad
        Q(z) = \frac{4\left[x_h^2\Omega^2-V_\epsilon(z)\right]}{f^2(z)}
    \end{equation}
with $V_\epsilon(z)$ defined in (\ref{ourODE}). The asymptotic behaviour of the solutions to equation (\ref{ODEas}) can be obtained by the method outlined in \cite{Olver1994MAA}. For this purpose, we start by observing that
\begin{equation}
    P(z) = \sum_{\kappa=0}^\infty\frac{\mathfrak{f}_\kappa}{z^k} = \mathcal{O}\left(\frac{1}{z^2}\right), \qquad
    Q(z) = \sum_{\kappa=0}^\infty\frac{\mathfrak{g}_\kappa}{z^k}=4x_h^2\Omega^2+\frac{8x_h\Omega^2}{z}+\mathcal{O}\left(\frac{1}{z^2}\right).
\end{equation}
Moreover, at least one of the coefficients $\mathfrak{f}_0$, $\mathfrak{g}_0$, $\mathfrak{g}_1$ is nonzero, and, therefore, a formal solution to (\ref{ODEas}) is represented by \cite{Olver1994MAA}
\begin{equation}\label{olvers}
    \psi^{(j)}_{\Omega\ell\epsilon}(z) = z^{\mu_j}e^{\lambda_j z}\sum_{\kappa=0}^\infty\frac{a_{\kappa,j}}{z^\kappa}, \qquad j \in \{1,2\},
\end{equation}
where $\lambda_1$, $\lambda_2$, $\mu_1$ and $\mu_2$ are the roots of the characteristic equations
\begin{equation}\label{chareqns}
   \lambda^2+\mathfrak{f}_0\lambda+\mathfrak{g}_0=0,\quad
   \mu_j=-\frac{\mathfrak{f}_1\lambda_j+\mathfrak{g}_1}{\mathfrak{f}_0+2\lambda_j}.
\end{equation}
A straightforward computation indicates that $\lambda_\pm = \pm 2ix_h\Omega$ and $\mu_\pm = \pm 2i\Omega$. As a result, the QNM boundary condition at space-like infinity can be expressed as
\begin{equation}\label{QNMBCzinf}
    \psi_{\Omega\ell\epsilon}\underset{{z\to +\infty}}{\longrightarrow} z^{2i\Omega}e^{2i x_h\Omega z}.
\end{equation}
\end{enumerate}
At this point, we can transform the radial function $\psi_{\Omega\ell\epsilon}(z)$ into a new radial function $\Phi_{\Omega\ell\epsilon}(z)$ such that the QNM boundary conditions are automatically implemented and  $\Phi_{\sigma\ell\epsilon}(z)$ is regular at $z = 1$ and at space-like infinity. To this aim, we consider the transformation
\begin{equation}\label{Ansatz}
    \psi_{\Omega\ell\epsilon}(z) = z^{2i(1+x_h\widetilde{\alpha})\Omega}(z-1)^{-2i x_h\widetilde{\alpha}\Omega}e^{2ix_h \Omega(z-1)} \Phi_{\Omega\ell\epsilon}(z)
\end{equation}
with $\widetilde{\alpha}$ defined as in (\ref{alphatilde}). After substitution of  (\ref{Ansatz}) into (\ref{ourODE}), we get the following ordinary differential equation for the radial eigenfunctions, namely
\begin{equation}\label{ODEznone}
    P_2(z)\Phi^{''}_{\Omega\ell\epsilon}(z) + P_1(z)\Phi^{'}_{\Omega\ell\epsilon}(z) + P_0(z)\Phi_{\Omega\ell\epsilon}(z) = 0
\end{equation}
with
\begin{eqnarray}
    P_2(z)&=&\frac{z^2(z-1)^2}{4}f^2(z),\\
    P_1(z)&=&z(z-1)f(z)\left\{\frac{z(z-1)}{4}f^{'}(z)-i\Omega f(z)\left[1+x_h\widetilde{\alpha}+(x_h-1)z-x_h z^2\right]\right\},\\
    P_0(z)&=&-\Omega^2 Q_+(z)Q_{-}(z)+i\Omega f(z)L(z)-z^2(z-1)^2 V_\epsilon(z),\\
    Q_\pm(z)&=&x_h z(z-1)\left[f(z)\pm 1\right]-f(z)\left(1+x_h\widetilde{\alpha}-z\right),\\
    L(z)&=&\frac{(z-1)^2}{2}\left[z(1+x_h z)f^{'}(z)-f(z)\right]-\frac{x_h\widetilde{\alpha}}{2}\left[z(z-1)f^{'}(z)+(1-2z)f(z)\right].
\end{eqnarray}
Let us now introduce the transformation $z = 2/(1-y)$, sending the point at infinity and the event horizon to $y = 1$ and $y = -1$, respectively. Moreover, a dot denotes differentiation with respect to the new variable $y$. Then, equation (\ref{ODEznone}) becomes
\begin{equation}\label{ODEnone}
    S_2(y)\ddot{\Phi}_{\Omega\ell\epsilon}(y) + S_1(y)\dot{\Phi}_{\Omega\ell\epsilon}(y) + S_0(y)\Phi_{\Omega\ell\epsilon}(y) = 0,
\end{equation}
where
\begin{eqnarray}
  S_2(y) &=& \frac{(1+y)^2}{4} f^2(y), \label{S2none} \\
  S_1(y) &=& i\Omega\frac{1+y}{(1-y)^2}f^2(y)\left[(1+y)(1+2x_h-y)-x_h\widetilde{\alpha}(1-y)^2\right]-\frac{(1+y)^2}{2(1-y)}f^2(y)+\frac{(1+y)^2}{4} f(y)\dot{f}(y), \label{S1none}\\
  S_0(y) &=& \Omega^2\Sigma_2(y)+i\Omega\Sigma_1(y)+\Sigma_0(y) \label{S0none}
\end{eqnarray}
with
\begin{eqnarray}
    \Sigma_2(y) &=& \frac{4x_h^2(1+y)^2}{(1-y)^4}-\frac{f^2(y)}{(1-y)^4}\left[(1+y)(1+2x_h-y)-x_h\widetilde{\alpha}(1-y)^2\right]^2,\\
    \Sigma_1(y) &=& \frac{f(y)}{2}\left\{
    \left(\frac{1+y}{1-y}\right)^2\left[(1+2x_h-y)\dot{f}(y)-f(y)\right]+\frac{x_h\widetilde{\alpha}}{1-y}\left[(3+y)f(y)-(1-y^2)\dot{f}(y)\right]
    \right\},\\
    \Sigma_0(y) &=& -\frac{4(1+y)^2}{(1-y)^4}V_\epsilon(y).
\end{eqnarray}
Notice that we must also require that $\Phi_{\Omega\ell\epsilon}(y)$ is regular at $y=\pm 1$. As a result of the transformation introduced above, we have 
\begin{eqnarray}
  f(y) &=& \frac{2x_h-1+y}{2x_h}\nonumber\\
  &+&e^{-\frac{2\alpha x_h}{1-y}}\left\{
    \frac{1}{2}\left[\frac{1-y}{x_h}+\alpha(1+q^2)\right]\cos{\left(\frac{2\alpha q x_h}{1-y}\right)+
    \frac{1}{2q}\left[\frac{(1-q^2)(1-y)}{2x_h}+\alpha(1+q^2)\right]\sin{\left(\frac{2\alpha q x_h}{1-y}\right)}}
  \right\}, \label{fv1}\\ 
  V_\epsilon(y) &=& \frac{(1-y)^2}{16}f(y)\left[\epsilon (1-y)\dot{f}(y)+\ell(\ell+1)\right].\label{fv2}
\end{eqnarray}
\begin{table}
\caption{Classification of the points $y=\pm 1$ for the relevant functions defined by  (\ref{S2none}), (\ref{S1none}), (\ref{S0none}), (\ref{fv1}), and (\ref{fv2}). The abbreviations $z$ ord $n$ and $p$ ord $m$ stand for zero of order $n$ and pole of order $m$, respectively.}
\begin{center}
\begin{tabular}{|c|c|c|c|c|c|c|c|}
\hline
$y$  & $f(y)$  & $V_\epsilon(y)$ & $S_2(y)$ & $S_1(y)$ & $S_0(y)$\\ \hline
$-1$ & z \mbox{ord} 1 & z \mbox{ord} 1 & z \mbox{ord} 4& z \mbox{ord} 3 & z \mbox{ord} 3 \\ \hline
$+1$ & $+1$  & z \mbox{ord} 2 & $+1$ & p \mbox{ord} 2 & p \mbox{ord} 2\\ \hline
\end{tabular}
\label{tableEinsnone}
\end{center}
\end{table}
Table~\ref{tableEinsnone} shows that the coefficients of the differential equation (\ref{ODEnone}) share a common zero of order $3$ at $y = -1$ while $y = 1$ is a pole of order $2$ for the coefficients $S_1$ and $S_0$. Hence, in order to apply the spectral method, we need to multiply (\ref{ODEnone}) by $(1-y)^2/(1+y)^3$. As a result, we end up with the following differential equation
\begin{equation}\label{ODEynonemod}
    M_2(y)\ddot{\Phi}_{\Omega\ell\epsilon}(y) + M_1(y)\dot{\Phi}_{\Omega\ell\epsilon}(y) + M_0(y)\Phi_{\Omega\ell\epsilon}(y) = 0,
\end{equation}
where
\begin{equation}\label{S210honone}
  M_2(y) = \frac{(1-y)^2}{4(1+y)}f^2(y), \qquad
  M_1(y) = i\Omega N_1(y)+N_0(y), \qquad
  M_0(y) = \Omega^2 C_2(y)+i\Omega C_1(y)+C_0(y)
\end{equation}
with
\begin{eqnarray}
    N_1(y) &=& f^2(y)\left[\frac{1+2x_h-y}{1+y}-x_h\widetilde{\alpha}\left(\frac{1-y}{1+y}\right)^2\right], \quad
    N_0(y) = \frac{f(y)}{4(1+y)}\frac{d}{dy}\left((1-y)^2 f(y)\right),\label{N0}\\
    C_2(y) &=& \frac{4x_h^2}{(1+y)(1-y)^2}-\frac{f^2(y)}{(1+y)^3(1-y)^2}\left[(1+y)(1+2x_h-y)-x_h\widetilde{\alpha}(1-y)^2\right]^2,\label{C2}\\
    C_1(y) &=& \frac{f(y)}{2(1+y)}\left\{(1+2x_h-y)\dot{f}(y)-f(y)+x_h\widetilde{\alpha}\frac{1-y}{(1+y)^2}\left[(3+y)f(y)-(1-y^2)\dot{f}(y)\right]\right\},\label{C1}\\
    C_0(y) &=& -\frac{4V_\epsilon(y)}{(1+y)(1-y)^2}.\label{C0}
\end{eqnarray}
It can be easily verified with \textsc{Maple} that
\begin{eqnarray}
    &&\lim_{y\to 1^{-}}M_2(y)=0=\lim_{y\to -1^{+}}M_2(y),\\
    &&\lim_{y\to 1^{-}}M_1(y)=ix_h\Omega,\quad
    \lim_{y\to -1^{+}}M_1(y)=i\Omega\Lambda_1+\Lambda_0,\\
    &&\lim_{y\to 1^{-}}M_0(y)=A_2\Omega^2 +A_0,\quad
     \lim_{y\to -1^{+}}M_0(y)=B_2\Omega^2+i\Omega B_1+B_0,
\end{eqnarray}
where the corresponding formulae for $\Lambda_0$, $\Lambda_1$, $A_0$, $A_2$, $B_0$, $B_1$, and $B_2$ can be found in the Appendix~\ref{AppendixA}. In the final step leading to the application of the spectral method, we recast the differential equation (\ref{ODEynonemod}) into the following form
\begin{equation}\label{TSCH}
  \widehat{L}_0\left[\Phi_{\Omega\ell\epsilon}, \dot{\Phi}_{\Omega\ell\epsilon}, \ddot{\Phi}_{\Omega\ell\epsilon}\right] +  i\widehat{L}_1\left[\Phi_{\Omega\ell\epsilon}, \dot{\Phi}_{\Omega\ell\epsilon}, \ddot{\Phi}_{\Omega\ell\epsilon}\right]\Omega +  \widehat{L}_2\left[\Phi_{\Omega\ell\epsilon}, \dot{\Phi}_{\Omega\ell\epsilon}, \ddot{\Phi}_{\Omega\ell\epsilon}\right]\Omega^2 = 0
\end{equation}
with
\begin{eqnarray}
  \widehat{L}_0\left[\Phi_{\Omega\ell\epsilon}, \dot{\Phi}_{\Omega\ell\epsilon}, \ddot{\Phi}_{\Omega\ell\epsilon}\right] &=& \widehat{L}_{00}(y)\Phi_{\Omega\ell\epsilon} + \widehat{L}_{01}(y)\dot{\Phi}_{\Omega\ell\epsilon} + \widehat{L}_{02}(y)\ddot{\Phi}_{\Omega\ell\epsilon},\label{L0none}\\
  \widehat{L}_1\left[\Phi_{\Omega\ell\epsilon}, \dot{\Phi}_{\Omega\ell\epsilon}, \ddot{\Phi}_{\Omega\ell\epsilon}\right] &=& \widehat{L}_{10}(y)\Phi_{\Omega\ell\epsilon} + \widehat{L}_{11}(y)\dot{\Phi}_{\Omega\ell\epsilon} + \widehat{L}_{12}(y)\ddot{\Phi}_{\Omega\ell\epsilon}, \label{L1none}\\
  \widehat{L}_2\left[\Phi_{\Omega\ell\epsilon}, \dot{\Phi}_{\Omega\ell\epsilon}, \ddot{\Phi}_{\Omega\ell\epsilon}\right] &=& \widehat{L}_{20}(y)\Phi_{\Omega\ell\epsilon} + \widehat{L}_{21}(y)\dot{\Phi}_{\Omega\ell\epsilon} + \widehat{L}_{22}(y)\ddot{\Phi}_{\Omega\ell\epsilon}.\label{L2none}
\end{eqnarray}
Moreover, in Table~\ref{tableNONEXT}, we have summarized the $\widehat{L}_{ij}$ appearing in (\ref{L0none})--(\ref{L2none}) and their limiting values at $y = \pm 1$.

\begin{table}
\caption{Definitions of the coefficients $\widehat{L}_{ij}$ and their corresponding behaviours at the endpoints of the interval $-1 \leq y \leq 1$. The symbols appearing in this table have been defined in the Appendix~\ref{AppendixA}.}
\begin{center}
\begin{tabular}{|c|c|c|c|c|c|c|c|}
\hline
$(i,j)$  & $\displaystyle{\lim_{y\to -1^+}}\widehat{L}_{ij}$  & $\widehat{L}_{ij}$ & $\displaystyle{\lim_{y\to 1^-}}\widehat{L}_{ij}$  \\ \hline
$(0,0)$ &  $B_0$          & $C_0$                  & $A_0$\\ \hline
$(0,1)$ &  $\Lambda_0$    & $N_0$                  & $0$\\ \hline
$(0,2)$ &  $0$            & $M_2$                  & $0$\\ \hline 
$(1,0)$ &  $B_1$          & $C_1$                  & $0$\\ \hline 
$(1,1)$ &  $\Lambda_1$    & $N_1$                  & $x_h$\\ \hline 
$(1,2)$ &  $0$            & $0$                    & $0$\\ \hline 
$(2,0)$ &  $B_2$          & $C_2$                  & $A_2$\\ \hline
$(2,1)$ &  $0$            & $0$                    & $0$\\ \hline
$(2,2)$ &  $0$            & $0$                    & $0$\\ \hline
\end{tabular}
\label{tableNONEXT}
\end{center}
\end{table}
We conclude this subsection by noting that extreme Type A BHs show a simple zero at the event horizon in the function $f(z)$ due to root coalescence occurring behind it. This characteristic enables the QNM calculations for this BH category to be conducted similarly to those for non-extreme cases.

\subsection{The extreme case}
\label{Sec2b}

In this scenario characterized by $\alpha = \alpha_e$, we introduce the rescalings $x=r/(2M)$, and $\xi = x/x_e$. Hence, the $g_{00}$ metric coefficient becomes
\begin{equation}\label{fze}
    f_e(\xi)=1-\frac{h_e(\xi)}{x_e \xi},\quad
    h_e(\xi)=1-e^{-\alpha_e x_e \xi}\left\{
    \left[1+\frac{\alpha_e x_e}{2}(1+q^2)\xi\right]\cos{(\alpha_e x_e q\xi)+
    \frac{1}{2q}\left[1-q^2+\alpha_e x_e(1+q^2)\xi\right]\sin{(\alpha_e x_e \xi)}}\right\}.
\end{equation}
The radial part of a massless Klein-Gordon field can be expressed in the equivalent form
\begin{equation}\label{ourODEext}
    \frac{f_e(\xi)}{4}\frac{d}{d\xi}\left(f_e(\xi)\frac{d\psi_{\Omega\ell\epsilon}}{d\xi}\right)+\left[x_e^2\Omega^2-V_\epsilon(\xi)\right]\psi_{\Omega\ell\epsilon}(\xi)=0,\qquad
    V_\epsilon(\xi)=\frac{f_e(\xi)}{4}\left[\frac{\epsilon}{\xi}\frac{df_e}{d\xi}+\frac{\ell(\ell+1)}{\xi^2}\right],\qquad
    \Omega=M\omega
\end{equation} 
with $f_e(\xi)$ as given in (\ref{fze}). In the case of extreme BHs of type B, $f_e(\xi)$ exhibits a zero of order two at $\xi = 1$. This implies that $f_e(1) = 0 = f^{'}(1)$ where the prime denotes differentiation with respect to $\xi$. This observation allows us to derive the following functional relations
\begin{eqnarray}
\cos{(\alpha_e x_e q)}&=&-\frac{2e^{\alpha_e x_e}\left[\alpha_e^2 x_e(x_e-1)(q^2+1)^2+\alpha_e x_e (q^2+1)+1-q^2\right]}{x_e\alpha_e^2(q^2+1)^2(\alpha_e q^2 x_e+\alpha_e x_e+2)},\\
\sin{(\alpha_e x_e q)}&=& \frac{2qe^{\alpha_e x_e}}{\alpha_e^2 x_e(q^2+1)^2},
\end{eqnarray}
which, in turn, plays an important role in simplifying the forthcoming computations. In order to derive the QNM boundary conditions at the event horizon and at infinity, we first need to determine the asymptotic behaviour of the radial solution $\psi_{\Omega\ell\epsilon}$ as $\xi \to 1^{+}$ and as $\xi \to +\infty$. We can then accurately extract the QNM boundary conditions from this asymptotic data.
\begin{enumerate}
\item 
{\underline{Asymptotic behaviour as $\xi \to 1^+$}}:
Taking into account that $\xi = 1$ is a double zero of $f_e(\xi)$ in the case of extreme Type B BHs, we can represent the latter in the form $f_e(\xi) = (\xi-1)^2 h(\xi)$ where $h$ is an analytic function at $\xi = 1$ with the property that 
\begin{equation}
h(1) = \frac{f_e^{''}(1)}{2}=\frac{\alpha_e^3 x_e^2(x_e-1)(q^2+1)^2+2\alpha_e^2 x_e^2(q^2+1)+2\alpha_e x_e(1-q^2)-2}{2\left[\alpha_e x_e(q^2+1)+2\right]}. 
\end{equation}
Note that for all scenarios showcased in Figure~\ref{figure01}, we have $h(1)>0$. This representation enables us to reformulate (\ref{ourODEext}) in the form
    \begin{eqnarray}
        &&\frac{d^2\psi_{\Omega\ell\epsilon}}{d\xi^2}+\mathfrak{p}(\xi)\frac{d\psi_{\Omega\ell\epsilon}}{d\xi}+\mathfrak{q}(\xi)\psi_{\Omega\ell\epsilon}(\xi)=0,\label{ODEZe}\\
        &&\mathfrak{p}(\xi)=\frac{2}{\xi-1}+\frac{h^{'}(\xi)}{h(\xi)},\quad
        \mathfrak{q}(\xi)=\frac{4x_e^2\Omega^2}{(\xi-1)^4 h^2(\xi)}-\frac{1}{(\xi-1)^2 h(\xi)}\left[\frac{\epsilon}{\xi}f_e^{'}(\xi)+\frac{\ell(\ell+1)}{\xi^2}\right].
    \end{eqnarray}
Because $\mathfrak{q}$ has a fourth-order pole at $\xi = 1$, it follows that $\xi = 1$ qualifies as an irregular singularity, rendering Frobenius's theory inapplicable in this scenario. On the other hand, since for $k = 1$ we have
\begin{equation}\label{ranks}
    (\xi-1)^{k+1}\mathfrak{p}(\xi)=\mathcal{O}(\xi-1),\quad 
    (\xi-1)^{2k+2}\mathfrak{q}(\xi)=\mathfrak{d}_0+\mathcal{O}(\xi-1)^2,\quad \mathfrak{d}_0=\frac{4x_e^2\Omega^2}{h^2(1)}
\end{equation}
with $\mathfrak{d}_0 \neq 0$, then, according to \cite{Bender1999}, $\xi = 1$ is an irregular singular point of rank one. Consequently, the leading behaviour of the solutions to equation (\ref{ODEZe}) in a neighbourhood of the event horizon can be obtained using the method outlined in \cite{Olver1994MAA}. To this purpose, we start by observing that by means of the transformation $\tau = (\xi-1)^{-1}$, (\ref{ODEZe}) becomes
\begin{eqnarray}
    &&\frac{d^2\psi_{\Omega\ell\epsilon}}{d\tau^2}+\mathfrak{C}(\tau)\frac{d\psi_{\Omega\ell\epsilon}}{d\tau}+\mathfrak{D}(\tau)\psi_{\Omega\ell\epsilon}(\tau)=0,\\
    &&\mathfrak{C}(\tau)=\sum_{\kappa=0}^\infty\frac{\mathfrak{c}_\kappa}{\tau^k}=\mathcal{O}\left(\frac{1}{\tau^2}\right),\quad
    \mathfrak{D}(\tau)=\sum_{\kappa=0}^\infty\frac{\mathfrak{d}_\kappa}{\tau^k}=\mathfrak{d}_0+\frac{\mathfrak{d}_1}{\tau}+\mathcal{O}\left(\frac{1}{\tau^2}\right)
\end{eqnarray}
with $\mathfrak{d}_0$ given by (\ref{ranks}) and 
\begin{equation}
\mathfrak{d}_1=\frac{4x_e^2\Omega^2\left[2\alpha_e^4 x_e^3(x_e-1)(q^2+1)^2-\alpha_e^3 x_e^2(q^2+1)(q^2-4x_e+1)-2\alpha_e^2 x_e^2(q^2-3)-4\alpha_e x_e q^2 -6\right]}{3\left[\alpha_e x_e(q^2+1)+2\right]h^3(1)}
\end{equation}
Since at least one of the coefficients $\mathfrak{c}_0$, $\mathfrak{d}_0$, $\mathfrak{d}_1$ is nonzero, a formal solution to (\ref{ODEZe}) is given by \cite{Olver1994MAA}
\begin{equation}
    \psi^{(\pm)}_{\Omega\ell\epsilon}(\tau)=\tau^{\mu_\pm}e^{\lambda_\pm \tau}\sum_{\kappa=0}^\infty\frac{\mathfrak{a}_{\kappa,\pm}}{\tau^\kappa},
\end{equation}
where $\lambda_\pm$, and $\mu_\pm$ are the roots of the characteristic equations
\begin{equation}
   \lambda_\pm^2+\mathfrak{d}_0=0,\quad
   \mu_\pm=-\frac{\mathfrak{d}_1}{2\lambda_\pm}.
\end{equation}
A straightforward computation shows that
\begin{eqnarray}
    \lambda_\pm&=&\pm\frac{2i x_e\Omega}{h(1)},\label{lmu1}\\
    \mu_\pm&=&\pm\frac{i x_e\Omega\left[2\alpha_e^4 x_e^3(x_e-1)(q^2+1)^2-\alpha_e^3 x_e^2(q^2+1)(q^2-4x_e+1)-2\alpha_e^2 x_e^2(q^2-3)-4\alpha_e x_e q^2 -6\right]}{3\left[\alpha_e x_e(q^2+1)+2\right]h^2(1)}\label{lmu2}.
\end{eqnarray}
Since a radial field exhibiting solely inward radiation near the event horizon ($\xi \to 1^+$) corresponds, under the transformation $\tau = (\xi-1)^{-1}$, to an outward radiating field as we approach spatial infinity ($\tau \to +\infty^{-}$), it is necessary to choose the plus sign in the formulas above. Consequently, the correct QNM boundary condition at $\xi = 1$ reads
\begin{equation}\label{QNMBCe1}
    \psi_{\Omega\ell\epsilon}\underset{{\xi\to 1^+}}{\longrightarrow} (\xi-1)^{-\mu_+}\mbox{exp}\left(\frac{\lambda_+}{\xi-1}\right)
\end{equation}
with $\mu_+$ and $\lambda_+$ defined in (\ref{lmu1}) and (\ref{lmu2}).
\item 
{\underline{Asymptotic behaviour as $\xi \to +\infty$}}: By means of the transformation $\eta = 1/\xi$, we can verify that the point at infinity is again an irregular singular point of rank one. Therefore, the asymptotic behaviour of the solutions to equation (\ref{ODEZe}) can be derived according to the method outlined in \cite{Olver1994MAA}. To this purpose, we  observe that
\begin{equation}
    \mathfrak{p}(\xi)=\sum_{\kappa=0}^\infty\frac{\widehat{\mathfrak{f}}_\kappa}{\xi^k}=\mathcal{O}\left(\frac{1}{\xi^2}\right),\quad
    \mathfrak{q}(\xi)=\sum_{\kappa=0}^\infty\frac{\widehat{\mathfrak{g}}_\kappa}{\xi^k}=4x_e^2\Omega^2+\frac{8x_e\Omega^2}{\xi}+\mathcal{O}\left(\frac{1}{\xi^2}\right).
\end{equation}
With the help of (\ref{chareqns}), we find that $\lambda_\pm = \pm 2i x_e\Omega$ and $\mu_\pm = \pm 2i\Omega$. Hence, the QNM boundary condition at space-like infinity can be expressed as
 \begin{equation}\label{QNMBCzinfe}
    \psi_{\Omega\ell\epsilon}\underset{{\xi\to +\infty}}{\longrightarrow} \xi^{2i\Omega}e^{2i x_e\Omega\xi}.
    \end{equation}
\end{enumerate}
Let us transform the radial function $\psi_{\Omega\ell\epsilon}(\xi)$ into a new radial function $\Phi_{\Omega\ell\epsilon}(\xi)$ such that the QNM boundary conditions are automatically implemented and  $\Phi_{\sigma\ell s}(\xi)$ is regular at $\xi = 1$ and at space-like infinity. To this aim, we consider the transformation
\begin{equation}\label{Ansatze}
    \psi_{\Omega\ell\epsilon}(\xi)=\xi^{2i\Omega+\mu_+}(\xi-1)^{-\mu_+}e^{2ix_e \Omega(\xi-1)+\frac{\lambda_+}{\xi-1}} \Phi_{\Omega\ell\epsilon}(\xi).
\end{equation}
If we rewrite it in a more compact form, namely
\begin{eqnarray}
    \psi_{\Omega\ell\epsilon}(\xi)&=&\xi^{2ia\Omega}(\xi-1)^{-2i(a-1)\Omega}e^{2ix_e\Omega\eta(\xi)} \Phi_{\Omega\ell\epsilon}(\xi),\quad
    \eta(\xi)=\xi-1+\frac{1}{h(1)(\xi-1)},\\
    a&=&1+\frac{x_e\left[2\alpha_e^4 x_e^3(x_e-1)(q^2+1)^2-\alpha_e^3 x_e^2(q^2+1)(q^2-4x_e+1)-2\alpha_e^2 x_e^2(q^2-3)-4\alpha_e x_e q^2 -6\right]}{6\left[\alpha_e x_e(q^2+1)+2\right]h^2(1)},\label{a2}
\end{eqnarray}
and we replace it into (\ref{ODEZe}), we end up with the differential equation
\begin{equation}\label{ODEzext}
    P_{2e}(\xi)\Phi^{''}_{\Omega\ell\epsilon}(\xi)+P_{1e}(\xi)\Phi^{'}_{\Omega\ell\epsilon}(\xi)+P_{0e}(\xi)\Phi_{\Omega\ell\epsilon}(\xi)=0
\end{equation}
where
\begin{eqnarray}
    P_{2e}(\xi)&=&\frac{\xi^2(\xi-1)^2}{4}f^2_e(\xi),\\
    P_{1e}(\xi)&=&\xi(\xi-1)f_e(\xi)\left\{\frac{\xi(\xi-1)}{4}f^{'}_e(\xi)+i\Omega f_e(\xi)\left[x_e\xi(\xi-1)\eta^{'}(\xi)+\xi-a\right]\right\},\\
    P_{0e}(\xi)&=&-\mathfrak{Q}_+(\xi)\mathfrak{Q}_-(\xi)\Omega^2+i\Omega f_e(\xi)\mathfrak{L}(\xi)-\xi^2(\xi-1)^2 V_\epsilon(\xi),\\
    \mathfrak{Q}_\pm(\xi)&=&x_e \xi(\xi-1)f_e(\xi)\eta^{'}(\xi)\pm x_e\xi(\xi-1)+(\xi-a)f_e(\xi),\\
    \mathfrak{L}(\xi)&=&\frac{x_e}{2}\xi^2(\xi-1)^2\left[f_e(\xi)\eta^{'}(\xi)\right]^{'}+\frac{1}{2}\xi(\xi-1)(\xi-a)f_e^{'}(\xi)-\frac{1}{2}(\xi^2-2a\xi+a)f_e(\xi).
\end{eqnarray}
(The variable $a$ defined in Eq.~\eqref{a2} should not be mistaken for the one in~\eqref{Rel2}.)
Let us introduce the transformation $\xi = 2/(1-y)$ mapping the point at infinity and the event horizon to $y = 1$ and $y = -1$, respectively. Furthermore, a dot denotes differentiation with respect to the new variable $y$. Then, equation (\ref{ODEzext}) becomes
\begin{equation}\label{ODEye}
    S_{2e}(y)\ddot{\Phi}_{\Omega\ell\epsilon}(y)+S_{1e}(y)\dot{\Phi}_{\Omega\ell\epsilon}(y)+S_{0e}(y)\Phi_{\Omega\ell\epsilon}(y)=0,
\end{equation}
where
\begin{eqnarray}
    S_{2e}(y)&=&\frac{(1+y)^2}{4} f_e^2(y),\label{S2oe}\\
    S_{1e}(y)&=&i\Omega\frac{1+y}{1-y}f_e^2(y)\left[x_e(1-y^2)\dot{\eta}(y)+2-a(1-y)\right]-\frac{(1+y)^2}{2(1-y)}f_e^2(y)+\frac{(1+y)^2}{4}f_e(y)\dot{f}_e(y),\label{S1oe}\\
    S_{0e}(y)&=&\Omega^2\Sigma_{2e}(y)+i\Omega\Sigma_{1e}(y)+\Sigma_{0e}(y)\label{S0oe}
\end{eqnarray}
with
\begin{eqnarray}
\Sigma_{2e}(y)&=&\frac{4x_e^2(1+y)^2}{(1-y)^4}-\frac{f_e^2(y)}{(1-y)^2}\left[x_e(1-y^2)\dot{\eta}(y)+2-a(1-y)\right]^2,\\
\Sigma_{1e}(y)&=&\frac{x_e}{2}(1+y)^2 f_e(y)\dot{f}_e(y)\dot{\eta}(y)+\frac{x_e}{2}\frac{(1+y)^2}{1-y}f_e^2(y)\left[(1-y)\ddot{\eta}(y)-2\dot{\eta}(y)\right]+\nonumber\\
&&\frac{1+y}{2(1-y)}\left[2-a(1-y)\right]f_e(y)\dot{f}_e(y)-\frac{f_e^2(y)}{(1-y)^2}\left[2-2a(1-y)+\frac{a}{2}(1-y)^2\right],\\
\Sigma_{1e}(y)&=&-\frac{4(1+y)^2}{(1-y)^4}V_\epsilon(y).
\end{eqnarray}
and the requirement that $\Phi_{\Omega\ell\epsilon}(y)$ is regular at $y = \pm 1$. As a result of the transformation introduced above, we have 
\begin{eqnarray}
f(y) &=& \frac{2x_e-1+y}{2x_e}e^{-\frac{2\alpha_e x_e}{1-y}}\cdot\nonumber\\
     &&\left\{
    \frac{1}{2}\left[\frac{1-y}{x_e}+\alpha_e(1+q^2)\right]\cos{\left(\frac{2\alpha_e q x_e}{1-y}\right)+
    \frac{1}{2q}\left[\frac{(1-q^2)(1-y)}{2x_e}+\alpha(1+q^2)\right]\sin{\left(\frac{2\alpha_e q x_e}{1-y}\right)}}
    \right\}, \label{fe1}\\ 
V_\epsilon(y) &=& \frac{(1-y)^2}{16}f_e(y)\left[\epsilon (1-y)\dot{f}_e(y)+\ell(\ell+1)\right],\quad
\eta(y)=\frac{1+y}{1-y}+\frac{1-y}{h(1)(1+y)}.\label{fe2}
\end{eqnarray}
\begin{table}
\caption{Classification of the points $y = \pm 1$ for the relevant functions entering in (\ref{S2oe}), (\ref{S1oe}) and (\ref{S0oe}). The abbreviations $z$ ord $n$ and $p$ ord $m$ stand for zero of order $n$ and pole of order $m$, respectively.}
\begin{center}
\begin{tabular}{|l|l|l|l|l|l|l|l|}
\hline
$y$  & $f_e(y)$  & $V_\epsilon(y)$ & $\eta(y)$ & $S_{2e}(y)$ & $S_{1e}(y)$ & $S_{0e}(y)$\\ \hline
$-1$ & z \mbox{ord} 2 & z \mbox{ord} 2 & p \mbox{ord} 1 & z \mbox{ord} 6& z \mbox{ord} 4 & z \mbox{ord} 4 \\ \hline
$+1$ & $+1$  & z \mbox{ord} 2 & p \mbox{ord} 1 & $+1$ & p \mbox{ord} 2 & p \mbox{ord} 2\\ \hline
\end{tabular}
\label{table3}
\end{center}
\end{table}
Table~\ref{table3} shows that the coefficients of the  differential equation (\ref{ODEye}) share a common zero of order $4$ at $y = -1$ while $y = 1$ is a pole of order $2$ for the coefficients $S_{1e}(y)$ and $S_{0e}(y)$. Hence, in order to apply the spectral method, we need to multiply (\ref{ODEye}) by $(1-y)^2/(1+y)^4$. As a result, we end up with the following differential equation
\begin{equation}\label{ODEhynonee}
    M_{2e}(y)\ddot{\Phi}_{\Omega\ell\epsilon}(y)+M_{1e}(y)\dot{\Phi}_{\Omega\ell\epsilon}(y)+M_{0e}(y)\Phi_{\Omega\ell\epsilon}(y)=0,
\end{equation}
where
\begin{equation}\label{S210hononee}
  M_{2e}(y)=\frac{(1-y)^2}{4(1+y)^2}f_e^2(y),\quad
  M_{1e}(y)=i\Omega N_{1e}(y)+N_{0e}(y),\quad
  M_{0e}(y)=\Omega^2 C_{2e}(y)+i\Omega C_{1e}(y)+C_{0e}(y)
\end{equation}
with
\begin{eqnarray}
    N_{1e}(y)&=&\frac{1-y}{(1+y)^3}f_e^2(y)\left[x_e(1-y^2)\dot{\eta}(y)+2-a(1-y)\right],\quad
    N_{0e}(y)=\frac{f_e(y)}{4(1+y)^2}\frac{d}{dy}\left((1-y)^2 f_e(y)\right),\label{N0e}\\
    C_{2e}(y)&=&\frac{4x_e^2}{(1-y^2)^2}-\frac{f_e^2(y)}{(1+y)^4}\left[x_e(1-y^2)\dot{\eta}(y)+2-a(1-y)\right]^2,\label{C2e}\\
    C_{1e}(y)&=&\frac{x_e}{2}\left(\frac{1-y}{1+y}\right)^2 f_e(y)\dot{f}_e(y)\dot{\eta}(y)+\frac{x_e}{2}\frac{1-y}{(1+y)^2}f_e^2(y)\left[(1-y)\ddot{\eta}(y)-2\dot{\eta}(y)\right]+\nonumber\\
    &&\frac{1-y}{2(1+y)^3}\left[2-a(1-y)\right]f_e(y)\dot{f}_e(y)-\frac{f_e^2(y)}{(1+y)^4}\left[2-2a(1-y)+\frac{a}{2}(1-y)^2\right],\label{C1e}\\
    C_{0e}(y)&=&-\frac{4V_\epsilon(y)}{(1-y^2)^2}.\label{C0e}
\end{eqnarray}
It can be easily checked with \textsc{Maple} that
\begin{eqnarray}
    &&\lim_{y\to 1^{-}}M_{2e}(y)=0=\lim_{y\to -1^{+}}M_{2e}(y),\quad \lim_{y\to 1^{-}}M_{1e}(y)=\frac{1}{2}ix_e\Omega,\quad
    \lim_{y\to -1^{+}}M_{1e}(y)=i\Omega\Lambda_{1e},\\
    &&\lim_{y\to 1^{-}}M_{0e}(y)=A_{2e}\Omega^2 +A_{0e},\quad
     \lim_{y\to -1^{+}}M_{0e}(y)=B_{0e},
\end{eqnarray}
where the coefficients $\Lambda_{1e}$, $A_{2e}$, $A_{0e}$, $B_{2e}$, and $B_{0e}$ are given in Appendix~\ref{AppendixB}. Finally, in order to apply the spectral method, we rewrite the differential equation (\ref{ODEhynonee}) into the following form
\begin{equation}\label{TSCHe}
\widehat{L}^{(e)}_0\left[\Phi_{\Omega\ell\epsilon},\dot{\Phi}_{\Omega\ell\epsilon},\ddot{\Phi}_{\Omega\ell\epsilon}\right]+ i\widehat{L}^{(e)}_1\left[\Phi_{\Omega\ell\epsilon},\dot{\Phi}_{\Omega\ell\epsilon},\ddot{\Phi}_{\Omega\ell\epsilon}\right]\Omega+ \widehat{L}_2^{(e)}\left[\Phi_{\Omega\ell\epsilon},\dot{\Phi}_{\Omega\ell\epsilon},\ddot{\Phi}_{\Omega\ell\epsilon}\right]\Omega^2=0
\end{equation}
with
\begin{eqnarray}
\widehat{L}^{(e)}_0\left[\Phi_{\Omega\ell\epsilon},\dot{\Phi}_{\Omega\ell\epsilon},\ddot{\Phi}_{\Omega\ell\epsilon}\right]&=&\widehat{L}^{(e)}_{00}(y)\Phi_{\Omega\ell\epsilon}+\widehat{L}^{(e)}_{01}(y)\dot{\Phi}_{\Omega\ell\epsilon}+\widehat{L}^{(e)}_{02}(y)\ddot{\Phi}_{\Omega\ell\epsilon},\label{L0nonee}\\
\widehat{L}^{(e)}_1\left[\Phi_{\Omega\ell\epsilon},\dot{\Phi}_{\Omega\ell\epsilon},\ddot{\Phi}_{\Omega\ell\epsilon}\right]&=&\widehat{L}^{(e)}_{10}(y)\Phi_{\Omega\ell\epsilon}+\widehat{L}^{(e)}_{11}(y)\dot{\Phi}_{\Omega\ell\epsilon}+\widehat{L}^{(e)}_{12}(y)\ddot{\Phi}_{\Omega\ell\epsilon},\label{L1nonee}\\
\widehat{L}^{(e)}_2\left[\Phi_{\Omega\ell\epsilon},\dot{\Phi}_{\Omega\ell\epsilon},\ddot{\Phi}_{\Omega\ell\epsilon}\right]&=&\widehat{L}^{(e)}_{20}(y)\Phi_{\Omega\ell\epsilon}+\widehat{L}^{(e)}_{21}(y)\dot{\Phi}_{\Omega\ell\epsilon}+\widehat{L}^{(e)}_{22}(y)\ddot{\Phi}_{\Omega\ell\epsilon}.\label{L2nonee}
\end{eqnarray}
Moreover, in Table~\ref{tableEXT}, we have summarized the $\widehat{L}^{(e)}_{ij}$ appearing in (\ref{L0nonee})--(\ref{L2nonee}) and their limiting values at $y = \pm 1$.

\begin{table}
\caption{Definitions of the coefficients $\widehat{L}^{(e)}_{ij}$ and their corresponding behaviours at the endpoints of the interval $-1 \leq y \leq 1$. The symbols appearing in this table have been defined in the Appendix~\ref{AppendixB}.}
\begin{center}
\begin{tabular}{|l|l|l|l|l|l|l|l|}
\hline
$(i,j)$  & $\displaystyle{\lim_{y\to -1^+}}\widehat{L}^{(e)}_{ij}$  & $\widehat{L}^{(e)}_{ij}$ & $\displaystyle{\lim_{y\to 1^-}}\widehat{L}^{(e)}_{ij}$  \\ \hline
$(0,0)$ &  $B_{0e}$       & $C_{0e}$                  & $A_{0e}$\\ \hline
$(0,1)$ &  $0$            & $N_{0e}$                  & $0$\\ \hline
$(0,2)$ &  $0$            & $M_{2e}$                  & $0$\\ \hline 
$(1,0)$ &  $0$            & $C_{1e}$                  & $0$\\ \hline 
$(1,1)$ &  $\Lambda_{1e}$ & $N_{1e}$                  & $x_e/2$\\ \hline 
$(1,2)$ &  $0$            & $0$                       & $0$\\ \hline 
$(2,0)$ &  $0$       & $C_{2e}$                  & $A_{2e}$\\ \hline
$(2,1)$ &  $0$            & $0$                       & $0$\\ \hline
$(2,2)$ &  $0$            & $0$                       & $0$\\ \hline
\end{tabular}
\label{tableEXT}
\end{center}
\end{table}

\section{Numerical method and results}
\label{Sec3}

In order to solve the differential eigenvalue problem \eqref{TSCH} to determine the QNMs along with the corresponding frequencies $\Omega$, we have to discretise the differential operators $\widehat{L}_{j}[\cdot]$ with $j \in \{0,1,2\}$ defined in \eqref{L0none}-\eqref{L2none}. Since our problem is posed on the finite interval $[-1, 1]$ without any boundary conditions, more precisely, we only require that the QNMs be regular functions at $y = \pm 1$, then, it is natural to choose a Chebyshev-type spectral method \cite{Trefethen2000, Boyd2000}. Namely, we are going to expand the function $y \mapsto \Phi_{\Omega\ell\epsilon}(y)$ in the form of a truncated Chebyshev series
\begin{equation}\label{eq:exp}
  \Phi_{\Omega\ell\epsilon}(y)=\sum_{k=0}^{N} a_k T_k(y),
\end{equation}
where $N\ \in\ \mathbb{N}$ is kept as a numerical parameter, $\{a_k\}_{k=0}^{N}\ \subseteq\ \mathds{R}$, and $\{T_k(y)\}_{k=0}^{N}$ are the Chebyshev polynomials of the first kind
\begin{equation}
    T_k: [-1, 1]\ \longrightarrow\ [-1, 1]\,, \qquad y\ \longmapsto\ \cos\,\bigl(k\arccos y\bigr)\,.
\end{equation}
After substituting expansion \eqref{eq:exp} into the differential equation \eqref{TSCH}, we obtain an eigenvalue problem with polynomial coefficients. In order to translate it into the realm of numerical linear algebra, we employ the collocation method \cite{Boyd2000}. Specifically, rather than insisting that the polynomial function in \( y \) is identically zero (a condition equivalent to having polynomial solutions for the differential problem as per equation \eqref{TSCH}), we impose a weaker requirement. This involves ensuring that the polynomial vanishes at \( N+1 \) strategically selected points. The number $N+1$ coincides exactly with the number of unknown coefficients $\{a_k\}_{k=0}^{N}$. For the collocation points, we implemented the Chebyshev roots grid \cite{Fox1968}
\begin{equation}
  y_k = -\cos{\left(\frac{(2k+1)\pi}{2(n+1)}\right)},\quad k\in\{0, 1,\ldots,N\}.
\end{equation}
In our numerical codes, we also implemented the second option of the Chebyshev extrema grid
\begin{equation}
  y_k = -\cos{\left(\frac{k\pi}{n}\right)},\quad k\in\{0, 1,\ldots,N\}.
\end{equation}
The users are free to choose their favourite collocation points. Notice that we used the roots grid in our computation, and in any case, the theoretical performance of the two available options is known to be absolutely comparable \cite{Fox1968, Boyd2000}.

Upon implementing the collocation method, we derive a classical matrix-based quadratic eigenvalue problem, as detailed in \cite{Tisseur2001}
\begin{equation}\label{eq:eig}
  (M_0 + iM_1\Omega + M_2\Omega^2)\bf{a} =\bf{0}.
\end{equation}
In this formulation, the square real matrices $M_{j}$, each of size $(N+1)\times(N+1)$ for $j \in \{0,1,2\}$, represent the spectral discretizations of the operators $\widehat{L}^{(e)}_{j}[\cdot]$, respectively. The problem \eqref{eq:eig} is solved numerically with the \textsc{polyeig} function from \textsc{Matlab}. This polynomial eigenvalue problem yields \(2(N+1)\) potential values for the parameter \(\Omega\). To discern the physical values of \(\Omega\) that correspond to the BH's QNMs, we first overlap the root plots for various values of \(N\) in equation \eqref{eq:exp}, such as \(N \in \{200, 250, 300\}\). We then identify the consistent roots whose positions remain stable across these different \(N\) values.

In order to reduce the rounding and other floating point errors, we performed all our computations with multiple precision arithmetic that is built in \textsc{Maple} and which is brought into \textsc{Matlab} by the \textsc{Advanpix} toolbox \cite{mct2015}. All numerical computations reported in this study have been performed with $300$ decimal digits accuracy. This measure could be considered as an overkill. However, this is not at all the case. Namely, we performed comparisons with QNMs computed in the standard double-precision floating-point arithmetic (as specified in the \texttt{IEEE-754-2008} standard), and the obtained spectra were highly distorted and inaccurate beyond a few first QNMs. That is why we decided in our study to sacrifice the speed of our computations for the sake of the robustness of the reported values.

Tables~\ref{table:1} through \ref{table:15} present typical values of the QNMs for Lee-Wick BHs under scalar, electromagnetic, and gravitational perturbations based on different choices of the parameters $q$ and $\alpha$. This is the first study to report such findings. All numerical values of the QNMs listed in the tables have been rounded to the fifth significant digit.

The transition of the Lee-Wick BH into the Schwarzschild metric, as $\alpha$ increases (see equation (\ref{hxfinal})), serves as a foundation for validating our numerical method. By selecting a sufficiently large $\alpha$ value, computing the QNMs, and demonstrating their agreement with the results obtained by \cite{Leaver1985PRSLA, IYER1987PRD, Mamani2022EPJC} for the classical Schwarzschild BH in cases $s \in \{0, 1, 2\}$, we can substantiate the accuracy of our approach. Relevant comparisons are detailed in the fourth column of Tables~\ref{table:1} and \ref{table:2} for $s = 0$, Tables~\ref{table:4} and \ref{table:5} for $s = 1$, and Tables~\ref{table:7} and \ref{table:8} for $s = 2$. Moreover, Tables~\ref{table:10}, \ref{table:11}, and \ref{table:12} display the QNMs for extreme Lee-Wick BHs of type B. In this case, owing to the computational challenges associated with the extremal case and the limitations of the spectral method, fewer frequencies were calculated for larger values of $\alpha$. Finally, Tables~\ref{table:13}, \ref{table:14}, and \ref{table:15} present the QNMs for extreme  Lee-Wick BHs of type A, which can be calculated using the same formulation of the non-extreme situation, as discussed in Sec.~\ref{Sec2a}.

An interesting feature observed in the nearly extremal cases of type-B BHs presented in Tables \ref{table:2}, \ref{table:3}, \ref{table:5}, \ref{table:6},  \ref{table:8}, \ref{table:9}, as well as the extreme Lee-Wick BHs of type A (referenced in Tables~\ref{table:13}, \ref{table:14}, \ref{table:15}), is the emergence of purely imaginary QNMs. Since the occurrence of such modes is accentuated as one approaches the extremal configurations, they are expected to be present also for extreme BHs of type B (they were not calculated in Tables~\ref{table:10}, \ref{table:11}, and \ref{table:12}, however, due to the complexities involved). The presence of QNMs with purely negative imaginary parts is significant because these modes represent perturbations that decay exponentially over time without oscillating. This is characteristic of an overdamped system in classical mechanics, where the system returns to equilibrium as quickly as possible without any oscillatory behaviour. In the context of BHs, this implies that any perturbations dampen smoothly into the BH and might suggest a highly efficient mechanism for settling down after being perturbed, which can be indicative of very specific or unique conditions at the BH's horizon.

Moreover, these modes are similar in nature to the so-called near-extremal frequencies detected by \cite{Cardoso2018PRL}. On the other hand, they can be thought of as emerging from the so-called zero-damped modes, which are represented by a sequence of QNMs converging to a purely imaginary number in the extremal limit \cite{Joykutty2022AHP}. It is interesting to observe that such modes also appear in the nearly extremal regimes of Reissner-Nordstr\"{o}m \cite{Hod2010PLA, Hod2012PLB, Hod2015PLB}, Reissner-Nordstr\"{o}m-de Sitter \cite{Cardoso2018PRD, Destounis2019PLB, Destounis2019JHEP, Joykutty2022AHP}, Kerr \cite{Hod2008PRDa, Hod2009PRD, Hod2011PRD, Yang2013PRDa, Yang2013PRDb}, Kerr-Newman BHs \cite{Hod2008PLB, Dias2015PRL}, and black strings \cite{Wuthicharn2021IJMPD}.

\begin{table} [H]
\centering
\caption{This table details the quasinormal frequencies for scalar perturbations (\(s = 0\)) in the context of a non-extreme Lee-Wick BH, with $q = 0.1$ and $q = 0.5$ for different settings of the parameter $\alpha$. Particularly, the fourth and sixth columns highlight scenarios with large \(\alpha\) values, where, as expected, the QNMs get closer to the numerical outcomes obtained by \cite{Leaver1985PRSLA,Mamani2022EPJC} through the continued fraction method, specifically for a Schwarzschild BH (see third column).The fifth and seventh columns correspond to choices of the parameter $\alpha$ slightly above the corresponding value of $\alpha$ characterizing an extreme Lee-Wick BH. The numerical values were derived using the spectral method with \(200\) polynomials, which allowed to achieve an accuracy of \(200\) digits. In this context, \(\Omega\) signifies the dimensionless frequency, as introduced in equation (\ref{ourODE}). The round bracket $(q,\alpha)$ indicates the corresponding choice of the parameters $q$ and $\alpha$.}
\label{table:1}
 \vspace*{1em}
 \begin{tabular}{||c| c| c| c| c| c|c||} 
 \hline
 $\ell$ & 
 $n$    & 
 $\Omega_\text{Schwarzschild}$  \cite{Mamani2022EPJC} &
 $\Omega$, $(0.1,100)$ & 
 $\Omega$, $(0.1,5.1)$ & 
 $\Omega$, $(0.5,100)$ & 
 $\Omega$, $(0.5,4.0)$ \\ [0.5ex] 
 \hline\hline
 $0$ & $0$ &$0.1105-0.1049i$ &$0.1105-0.1049i$ & $0.1065-0.0856i$ & $0.1105-0.1049i$ & $0.1032-0.0887i$ \\ 
     & $1$ &$0.0861-0.3481i$ &$0.0861-0.3481i$ & $0.0441-0.3021i$ & $0.0861-0.3481i$ & $0.0310-0.3113i$ \\
\hline
 $1$ & $0$ &$0.2929-0.0977i$ &$0.2929-0.0977i$ & $0.2970-0.0794i$ & $0.2929-0.0977i$ & $0.2871-0.0854i$ \\
     & $1$ &$0.2645-0.3063i$ &$0.2645-0.3063i$ & $0.2560-0.2449i$ & $0.2645-0.3063i$ & $0.2376-0.2597i$ \\
     & $2$ &$0.2295-0.5401i$ &$0.2295-0.5401i$ & $0.0000-0.3645i$ & $0.2295-0.5401i$ & $0.1456-0.4537i$ \\ 
     & $3$ &$0.2033-0.7883i$ &$0.2033-0.7883i$ & $0.0000-0.6547i$ & $0.2033-0.7883i$ & $0.0000-0.4344i$ \\
\hline
 $2$ & $0$ &$0.4836-0.0968i$ &$0.4836-0.0968i$ & $0.4926-0.0789i$ & $0.4836-0.0968i$ & $0.4769-0.0858i$ \\
     & $1$ &$0.4639-0.2956i$ &$0.4639-0.2956i$ & $0.4672-0.2390i$ & $0.4639-0.2956i$ & $0.4449-0.2580i$ \\
     & $2$ &$0.4305-0.5086i$ &$0.4305-0.5086i$ & $0.4155-0.4060i$ & $0.4305-0.5086i$ & $0.3816-0.4303i$ \\
     & $3$ &$0.3939-0.7381i$ &$0.3939-0.7381i$ & $0.0000-0.3945i$ & $0.3939-0.7381i$ & $0.2934-0.6158i$ \\ [0.5ex] 
 \hline
 \end{tabular}
\end{table}

\begin{table} [H]
\centering
\caption{This table details the quasinormal frequencies for scalar perturbations (\(s = 0\)) in the context of a non-extreme Lee-Wick BH, with $q = 1$ and $q = 2$ for different settings of the parameter $\alpha$. Particularly, the third and fifth columns highlight scenarios with large \(\alpha\) values, where, as expected, the QNMs closely align with the numerical outcomes obtained by \cite{Leaver1985PRSLA,Mamani2022EPJC} through the continued fraction method, specifically for a Schwarzschild BH (see third column in Table~\ref{table:1}). The fourth and sixth columns correspond to choices of the parameter $\alpha$ slightly above the corresponding value of $\alpha$ characterizing an extreme Lee-Wick BH. \textbf{The seventh and eighth columns correspond to BHs with two horizons, with $\alpha$ between two cases of extreme horizons of type B (see Table~\ref{tableEins}).} The numerical values were derived using the spectral method with \(200\) polynomials, which allowed to achieve an accuracy of \(200\) digits. In this context, \(\Omega\) signifies the dimensionless frequency, as introduced in equation (\ref{ourODE}). The round bracket $(q,\alpha)$ indicates the corresponding choice of the parameters $q$ and $\alpha$.  The QNMs appearing in the first and third lines of the fourth column look identical to the first five digits, but they are not the same because for $\ell=0$ and $\ell=1$, the elements in the matrix $M_0$ (see equation (\ref{eq:eig})) differ only beyond the first 100 digits.}
\label{table:2}
 \vspace*{1em}
 \begin{tabular}{||c|c|c|c|c|c|c|c||}
 \hline
 $\ell$ & 
 $n$    & 
 $\Omega$, $(1, 100)$ & 
 $\Omega$, $(1, 2.2)$ & 
 $\Omega$, $(2, 100)$ & 
 $\Omega$, $(2, 0.7)$ &
 $\Omega$, $(2, 1.0)$ &
 $\Omega$, $(2, 2.0)$
 \\ [0.5ex] 
 \hline\hline
 $0$ & $0$ & $0.1105-0.1049i$ & $0.0863-0.0917i$ & $0.1105-0.1049i$ & $0.0000-0.0457i$ & $0.0000-0.6041i$ & $0.0000-3.9700i$ \\ 
     & $1$ & $0.0861-0.3481i$ & $0.0000-0.1593i$ & $0.0861-0.3481i$ & $0.0000-0.0910i$ & $0.0000-1.1985i$ & $0.0000-7.8641i$ \\
\hline
 $1$ & $0$ & $0.2929-0.0977i$ & $0.0863-0.0917i$ & $0.2929-0.0977i$ & $0.1567-0.0839i$ & $0.0000-0.7402i$ & $0.2756-0.1493i$ \\
     & $1$ & $0.2645-0.3063i$ & $0.0000-0.1593i$ & $0.2645-0.3063i$ & $0.0000-0.0718i$ & $0.0000-1.2862i$ & $0.0000-4.3324i$ \\
     & $2$ & $0.2295-0.5401i$ & $0.0000-0.2398i$ & $0.2295-0.5401i$ & $0.0000-0.1140i$ & $0.0000-1.8453i$ & $0.0000-8.0647i$ \\ 
     & $3$ & $0.2033-0.7883i$ & $0.0000-0.3203i$ & $0.2033-0.7883i$ & $0.0000-0.1567i$ & $0.0000-2.4092i$ & $0.0000-11.794i$ \\
\hline
 $2$ & $0$ & $0.4836-0.0968i$ & $0.4226-0.0954i$ & $0.4836-0.0968i$ & $0.2763-0.0967i$ & $0.0000-0.9601i$ & $0.4729-0.1712i$ \\
     & $1$ & $0.4639-0.2956i$ & $0.3568-0.2740i$ & $0.4639-0.2956i$ & $0.0000-0.1037i$ & $0.0000-1.4426i$ & $0.0000-4.9772i$ \\
     & $2$ & $0.4305-0.5086i$ & $0.0000-0.2071i$ & $0.4305-0.5086i$ & $0.0000-0.1453i$ & $0.0000-1.9623i$ & $0.0000-8.4428i$ \\
     & $3$ & $0.3939-0.7381i$ & $0.0000-0.2801i$ & $0.3939-0.7381i$ & $0.0000-0.1870i$ & $0.0000-2.5015i$ & $0.0000-12.055i$ \\ [0.5ex] 
 \hline
 \end{tabular}
\end{table}

\begin{table} [H]
\centering
\caption{This table details the quasinormal frequencies for scalar perturbations (\(s = 0\)) in the context of a non-extreme Lee-Wick BH, with $q = 2$ for different settings of the parameter $\alpha$. \textbf{The choices of the parameter $\alpha$ are slightly above the corresponding values characterizing extreme Lee-Wick BHs of Type B (fourth column) and A (fifth column), see Table~\ref{tableEins}.} The numerical values were derived using the spectral method with \(300\) polynomials, which allowed to achieve an accuracy of \(300\) digits. In this context, \(\Omega\) signifies the dimensionless frequency, as introduced in equation (\ref{ourODE}). The round bracket $(q,\alpha)$ indicates the corresponding choice of the parameters $q$ and $\alpha$. We observe a significant deviation in the quasinormal modes compared to those of a Schwarzschild black hole, as detailed in the third column of Table~\ref{table:1}. Moreover, the occurence of purely imaginary QNMs is accentuated for $\alpha=4.1$, i.e., close to the extremal case of Type B.}
\label{table:3}
 \vspace*{1em}
 \begin{tabular}{||c|c|c|c|c|c|c||} 
 \hline
 $\ell$ & 
 $n$    & 
 $\Omega$, $(2, 3.0)$ &
 $\Omega$, $(2, 4.1)$ & 
 $\Omega$, $(2, 5.2)$ &
 $\Omega$, $(2, 6.0)$ &
 $\Omega$, $(2, 10)$ \\ [0.5ex] 
 \hline\hline
 $0$ & $0$ & $0.1272-0.1120i$  & $0.0936-0.1111i$ & $0.1083-0.1158i$ & $0.1147-0.1053i$ & $0.1103-0.1045i$ \\ 
     & $1$ & $0.0000-9.9773i$  & $0.0000-0.1452i$ & $0.0000-1.3151i$ & $0.0000-2.2774i$ & $0.0000-1.8085i$ \\
\hline
 $1$ & $0$ & $0.3350 -0.1015i$ & $0.2784-0.1167i$ & $0.3012-0.1066i$ & $0.2988-0.0909i$ & $0.2926-0.0977i$ \\
     & $1$ & $1.1452 -0.5173i$ & $0.0000-0.3247i$ & $0.0000-1.4397i$ & $0.3110-0.3061i$ & $0.2617-0.3052i$ \\
     & $2$ & $3.0889 -1.0411i$ & $0.0000-0.4749i$ & $0.0000-2.8076i$ & $0.0000-2.4410i$ & $0.8273-1.8470i$ \\ 
     & $3$ & $0.0000-10.5563i$ & $0.0000-0.6262i$ & $0.0000-4.1917i$ & $0.0000-4.8148i$ & $0.0000-2.0058i$ \\
\hline
 $2$ & $0$ & $0.5333-0.0909i$  & $0.4733-0.1227i$ & $0.4975-0.1007i$ & $0.4849-0.0847i$ & $0.4835-0.0971i$ \\
     & $1$ & $0.8883-0.3611i$  & $0.0000-0.2202i$ & $0.4956-0.3338i$ & $0.5025-0.2702i$ & $0.4618-0.2966i$ \\
     & $2$ & $1.2183-0.4733i$  & $0.0000-0.3634i$ & $0.0000-1.6304i$ & $0.0000-2.7027i$ & $0.9850-1.8445i$ \\
     & $3$ & $3.1152-1.0281i$  & $0.0000-0.5095i$ & $0.0000-2.9448i$ & $0.0000-4.9950i$ & $0.0000-2.2981i$ \\ 
     & $4$ & $0.0000-11.6212i$ & $0.0000-0.6576i$ & $0.0000-4.2998i$ & $0.0000-7.3485i$ & $1.3992-3.1071i$ \\ [0.5ex] 
 \hline
 \end{tabular}
\end{table}

\begin{table} [H]
\centering
\caption{This table details the quasinormal frequencies for electromagnetic perturbations (\(s = 1\)) in the context of a non-extreme Lee-Wick BH, with $q = 0.1$ and $q = 0.5$ for different settings of the parameter $\alpha$. Particularly, the fourth and sixth columns highlight scenarios with large \(\alpha\) values, where, as expected, the QNMs get closer to the numerical values obtained by \cite{Leaver1985PRSLA,Mamani2022EPJC}, specifically for a Schwarzschild BH (see third column). The fifth and seventh columns correspond to choices of the parameter $\alpha$ slightly above the corresponding value of $\alpha$ characterizing an extreme Lee-Wick BH. The numerical values were derived using the spectral method with \(300\) polynomials, which allowed to achieve an accuracy of \(300\) digits. In this context, \(\Omega\) signifies the dimensionless frequency, as introduced in equation (\ref{ourODE}). The round bracket $(q,\alpha)$ indicates the corresponding choice of the parameters $q$ and $\alpha$.}
\label{table:4}
 \vspace*{1em}
 \begin{tabular}{||c|c|c|c|c|c|c|c||}
 \hline
 $\ell$ & 
 $n$    & 
$\Omega_\text{Schwarzschild}$ \cite{Mamani2022EPJC}&
 $\Omega$, $(0.1, 100)$ & 
 $\Omega$, $(0.1, 5.1)$ & 
 $\Omega$, $(0.5, 100)$ & 
 $\Omega$, $(0.5, 4.0)$ \\ [0.5ex] 
 \hline\hline
 $1$ & $0$ &$0.2483-0.0925i$ &$0.2483-0.0925i$ & $0.2587-0.0724i$ & $0.2483-0.0925i$ & $0.2470-0.0775i$ \\ 
     & $1$ &$0.2145-0.2937i$ &$0.2145-0.2937i$ & $0.2180-0.2237i$ & $0.2145-0.2937i$ & $0.1996-0.2367i$ \\
\hline
 $2$ & $0$ &$0.4576-0.0950i$ &$0.4576-0.0950i$ & $0.4693-0.0762i$ & $0.4576-0.0950i$ & $0.4524-0.0826i$ \\
     & $1$ &$0.4365-0.2907i$ &$0.4365-0.2907i$ & $0.4444-0.2310i$ & $0.4365-0.2907i$ & $0.4216-0.2490i$ \\
     & $2$ &$0.4012-0.5016i$ &$0.4012-0.5016i$ & $0.3935-0.3930i$ & $0.4012-0.5016i$ & $0.3587-0.4169i$ \\
\hline
 $3$ & $0$ &$0.6569-0.0956i$ &$0.6569-0.0956i$ & $0.6721-0.0774i$ & $0.6569-0.0956i$ & $0.6498-0.0843i$ \\
     & $1$ &$0.6417-0.2897i$ &$0.6417-0.2897i$ & $0.6541-0.2334i$ & $0.6417-0.2897i$ & $0.6270-0.2537i$ \\
     & $2$ &$0.6138-0.4921i$ &$0.6138-0.4921i$ & $0.6175-0.3929i$ & $0.6138-0.4921i$ & $0.5790-0.4234i$ \\
     & $3$ &$0.5779-0.7063i$ &$0.5779-0.7063i$ & $0.5610-0.5586i$ & $0.5779-0.7063i$ & $0.5131-0.5839i$ \\ 
\hline
 $4$ & $0$ &$0.8531-0.0959i$ &$0.8531-0.0959i$ & $0.8721-0.0779i$ & $0.8531-0.0959i$ & $0.8444-0.0851i$ \\
     & $1$ &$0.8413-0.2893i$ &$0.8413-0.2893i$ & $0.8581-0.2344i$ & $0.8413-0.2893i$ & $0.8264-0.2559i$ \\
     & $2$ &$0.8187-0.4878i$ &$0.8187-0.4878i$ & $0.8296-0.3931i$ & $0.8187-0.4878i$ & $0.7886-0.4277i$ \\
     & $3$ &$0.7877-0.6942i$ &$0.7877-0.6942i$ & $0.7860-0.5555i$ & $0.7877-0.6942i$ & $0.7253-0.5933i$ \\
     & $4$ &$0.7515-0.9102i$ &$0.7515-0.9102i$ & $0.0000-0.4260i$ & $0.7515-0.9102i$ & $0.6851-0.7258i$ \\ [0.5ex] 
 \hline
 \end{tabular}
\end{table}

\begin{table} [H]
\centering
\caption{This table details the quasinormal frequencies for electromagnetic perturbations (\(s = 1\)) in the context of a non-extreme Lee-Wick BH, with $q = 1$ and $q = 2$ for different settings of the parameter $\alpha$. Particularly, the third and fifth columns highlight scenarios with large \(\alpha\) values, where, as expected, the QNMs get closer to the numerical results obtained by \cite{Mamani2022EPJC} through the spectral method, specifically for a Schwarzschild BH (see third column in Table~\ref{table:4}). The fourth and sixth columns correspond to choices of the parameter $\alpha$ slightly above the corresponding value of $\alpha$ characterizing an extreme Lee-Wick BH. The numerical values were derived using the spectral method with \(300\) polynomials, which allowed to achieve an accuracy of \(300\) digits. In this context, \(\Omega\) signifies the dimensionless frequency, as introduced in equation (\ref{ourODE}). The round bracket $(q,\alpha)$ indicates the corresponding choice of the parameters $q$ and $\alpha$.}
\label{table:5}
 \vspace*{1em}
 \begin{tabular}{||c|c|c|c|c|c|c|c||} 
 \hline
 $\ell$ & 
 $n$    & 
 $\Omega$, $(1, 100)$ & 
 $\Omega$, $(1, 2.2)$ & 
 $\Omega$, $(2, 100)$ & 
 $\Omega$, $(2, 0.7)$ &
 $\Omega$, $(2, 1.0)$ &
 $\Omega$, $(2, 2.0)$ \\ [0.5ex] 
 \hline\hline
 $1$ & $0$ & $0.2483-0.0925i$ & $0.2101-0.0795i$ & $0.2483-0.0925i$  & $0.1265-0.0650i$ & $0.1459-0.1001i$ & $0.2186-0.1330i$ \\
     & $1$ & $0.2145-0.2937i$ & $0.0000-0.2162i$ & $0.2145-0.2937i$  & $0.0000-0.0724i$ & $0.0000-0.8372i$ & $0.0000-5.3549i$ \\
\hline
 $2$ & $0$ & $0.4576-0.0950i$ & $0.3965-0.0896i$ & $0.4576-0.0950i$  & $0.2558-0.0848i$ & $0.3085-0.1315i$ & $0.4389-0.1694i$ \\
     & $1$ & $0.4365-0.2907i$ & $0.3357-0.2630i$ & $0.4365-0.2907i$  & $0.0000-0.1040i$ & $0.0000-1.0298i$ & $0.0000-5.8625i$ \\
     & $2$ & $0.4012-0.5016i$ & $0.0000-0.2077i$ & $0.4012-0.5016i$  & $0.0000-0.1458i$ & $0.0000-1.5468i$ & $0.0000-9.5012i$ \\
\hline
 $3$ & $0$ & $0.6569-0.0956i$ & $0.5760-0.0935i$ & $0.6569-0.0956i$  & $0.3878-0.0934i$ & $0.4949-0.1670i$ & $0.6247-0.1879i$ \\
     & $1$ & $0.6417-0.2897i$ & $0.5273-0.2838i$ & $0.6417-0.2897i$  & $0.0000-0.1365i$ & $0.0000-1.2883i$ & $0.8694-0.3265i$ \\
     & $2$ & $0.6138-0.4921i$ & $0.0000-0.2758i$ & $0.6138-0.4921i$  & $0.0000-0.1782i$ & $0.0000-1.7407i$ & $0.0000-6.6041i$ \\
     & $3$ & $0.5779-0.7063i$ & $0.0000-0.3489i$ & $0.5779-0.7063i$  & $0.0000-0.2200i$ & $0.0000-2.2319i$ & $0.0000-9.9220i$ \\
\hline
 $4$ & $0$ & $0.8531-0.0959i$ & $0.7523-0.0952i$ & $0.8531-0.0959i$  & $0.5130-0.0944i$ & $0.6625-0.1578i$ & $0.7852-0.1882i$ \\
     & $1$ & $0.8413-0.2893i$ & $0.7188-0.2900i$ & $0.8413-0.2893i$  & $0.0000-0.1692i$ & $0.4839-0.1895i$ & $1.0667-0.2957i$ \\
     & $2$ & $0.8187-0.4878i$ & $0.5777-0.3894i$ & $0.8187-0.4878i$  & $0.0000-0.2108i$ & $0.0000-1.5866i$ & $0.0000-7.5728i$ \\
     & $3$ & $0.7877-0.6942i$ & $0.0000-0.3440i$ & $0.7877-0.6942i$  & $0.0000-0.2525i$ & $0.0000-2.0034i$ & $0.0000-10.4725i$ \\
     & $4$ & $0.7515-0.9102i$ & $0.0000-0.4169i$ & $0.7515-0.9102i$  & $0.0000-0.2944i$ & $0.0000-2.4313i$ & $0.0000-13.8547i$ \\ [0.5ex]
 \hline
 \end{tabular}
\end{table}

\begin{table} [H]
\centering
\caption{This table details the quasinormal frequencies for electromagnetic perturbations (\(s = 1\)) in the context of a non-extreme Lee-Wick BH, with $q = 2$ for different settings of the parameter $\alpha$. The choices of the parameter $\alpha$ are slightly above the corresponding values of $\alpha$, characterizing extreme Lee-Wick BHs of Type A and B. The numerical values were derived using the spectral method with \(300\) polynomials, which allowed to achieve an accuracy of \(300\) digits. In this context, \(\Omega\) signifies the dimensionless frequency, as introduced in equation (\ref{ourODE}). The round bracket $(q,\,\alpha)$ indicates the corresponding choice of the parameters $q$ and $\alpha$.  We observe a significant deviation in the quasinormal modes compared to those of a Schwarzschild black hole, as detailed in the third column of Table~\ref{table:4}. The occurence of purely imaginary QNMs is accentuated for $\alpha=4.1$, i.e., close to the extremal case of Type B (see Table~\ref{tableEins}).}
\label{table:6}
 \vspace*{1em}
 \begin{tabular}{||c|c|c|c|c|c|c||} 
 \hline
 $\ell$ &
 $n$    &
 $\Omega$, $(2, 3.0)$ &
 $\Omega$, $(2, 4.1)$ &
 $\Omega$, $(2, 5.2)$ &
 $\Omega$, $(2, 6.0)$ &
 $\Omega$, $(2, 10)$ \\ [0.5ex] 
 \hline\hline
 $1$ & $0$ & $0.2831-0.1087i$  & $0.2284-0.1052i$ & $0.2515-0.1037i$ & $0.2555-0.0894i$ & $0.2480-0.0924i$ \\
     & $1$ & $0.0000-13.3904i$ & $0.0000-0.1786i$ & $0.0000-1.5319i$ & $0.0000-2.6360i$ & $0.2127-0.2918i$ \\
\hline
 $2$ & $0$ & $0.5064-0.0985i$  & $0.4432-0.1196i$ & $0.4703-0.1015i$ & $0.4612-0.0839i$ & $0.4574-0.0953i$ \\
     & $1$ & $0.0000-14.1949i$ & $0.0000-0.2208i$ & $0.4635-0.3321i$ & $0.4758-0.2724i$ & $0.4344-0.2912i$ \\
     & $2$ & $0.0000-23.5519i$ & $0.0000-0.3649i$ & $0.0000-1.6919i$ & $0.0000-2.8441i$ & $1.0305-1.8319i$ \\
\hline
 $3$ & $0$ & $0.6985-0.0849i$  & $0.6479-0.1258i$ & $0.6735-0.0965i$ & $0.6536-0.0817i$ & $0.6569-0.0959i$ \\
     & $1$ & $1.0142-0.2717i$  & $0.5947-0.3614i$ & $0.6807-0.3193i$ & $0.6731-0.2517i$ & $0.6406-0.2913i$ \\
     & $2$ & $1.2669-0.4453i$  & $0.0000-0.2691i$ & $0.0000-1.9036i$ & $0.6963-0.4651i$ & $0.6074-0.4936i$ \\
     & $3$ & $0.0000-15.3821i$ & $0.0000-0.4098i$ & $0.0000-3.1935i$ & $0.0000-3.1276i$ & $0.8117-0.8509i$ \\
\hline
 $4$ & $0$ & $0.8791-0.0720i$  & $0.8485-0.1297i$ & $0.8714-0.0918i$ & $0.8429-0.0828i$ & $0.8533-0.0961i$ \\
     & $1$ & $1.2047-0.2371i$  & $0.8066-0.3654i$ & $0.8887-0.3090i$ & $0.8611-0.2368i$ & $0.8412-0.2912i$ \\
     & $2$ & $1.4283-0.4021i$  & $0.0000-0.3201i$ & $0.0000-2.1511i$ & $0.8958-0.4353i$ & $0.8147-0.4921i$ \\
     & $3$ & $0.0000-16.9437i$ & $0.0000-0.4589i$ & $0.0000-3.3831i$ & $0.0000-3.4672i$ & $0.9566-0.9548i$ \\
     & $4$ & $0.0000-25.0259i$ & $0.0000-0.5999i$ & $0.0000-4.6903i$ & $0.0000-5.6405i$ & $1.2932-1.9074i$ \\[0.5ex]
 \hline
 \end{tabular}
\end{table}

\begin{table} [H]
\centering
\caption{This table details the quasinormal frequencies for gravitational perturbations (\(s = 2\)) in the context of a non-extreme Lee-Wick BH, with $q = 0.1$ and $q = 0.5$ for different settings of the parameter $\alpha$. Particularly, the fourth and sixth columns highlight scenarios with large \(\alpha\) values, where, as expected, the QNMs get closer to the numerical values obtained by \cite{Leaver1985PRSLA,Mamani2022EPJC}, specifically for a Schwarzschild BH (see third column). The fifth and seventh columns correspond to choices of the parameter $\alpha$ slightly above the corresponding value of $\alpha$ characterizing an extreme Lee-Wick BH. The numerical values were derived using the spectral method with \(300\) polynomials, which allowed to achieve an accuracy of \(300\) digits. In this context, \(\Omega\) signifies the dimensionless frequency, as introduced in equation (\ref{ourODE}). The round bracket $(q,\,\alpha)$ indicates the corresponding choice of the parameters $q$ and $\alpha$.}
\label{table:7}
 \vspace*{1em}
 \begin{tabular}{||c|c|c|c|c|c|c||}
 \hline
 $\ell$ & 
 $n$    & 
 $\Omega_\text{Schwarzschild}$ \cite{Mamani2022EPJC}&
 $\Omega$, $(0.1, 100)$ & 
 $\Omega$, $(0.1, 5.1)$ & 
 $\Omega$, $(0.5, 100)$ & 
 $\Omega$, $(0.5, 4.0)$ \\ [0.5ex]
 \hline\hline
 $2$ & $0$ & $0.3737-0.0890i$  &$0.3737-0.0890i$ & $0.4693-0.0762i$ & $0.3737-0.0890i$ & $0.3729-0.0698i$ \\ 
     & $1$ & $0.3467-0.2739i$  &$0.3467-0.2739i$ & $0.4444-0.2310i$ & $0.3467-0.2739i$ & $0.3508-0.2136i$ \\
     & $2$ & $0.3011-0.4783i$  &$0.3011-0.4783i$ & $0.3935-0.3930i$ & $0.3011-0.4783i$ & $0.2987-0.3667i$ \\
\hline
 $3$ & $0$ & $0.5994-0.0927i$  &$0.5994-0.0927i$ & $0.6196-0.0722i$ & $0.5994-0.0927i$ & $0.5945-0.0784i$ \\
     & $1$ & $0.5826-0.2813i$  &$0.5826-0.2813i$ & $0.6035-0.2182i$ & $0.5826-0.2813i$ & $0.5746-0.2368i$ \\
     & $2$ & $0.5517-0.4791i$  &$0.5517-0.4791i$ & $0.5701-0.3692i$ & $0.5517-0.4791i$ & $0.5300-0.3986i$ \\ 
     & $3$ & $0.5112-0.6903i$  &$0.5112-0.6903i$ & $0.5165-0.5287i$ & $0.5112-0.6903i$ & $0.4674-0.5461i$ \\
\hline
 $4$ & $0$ & $0.8092-0.0942i$  &$0.8092-0.0942i$ & $0.8319-0.0749i$ & $0.8092-0.0942i$ & $0.8019-0.0816i$ \\
     & $1$ & $0.7966-0.2843i$  &$0.7966-0.2843i$ & $0.8186-0.2256i$ & $0.7966-0.2843i$ & $0.7853-0.2458i$ \\
     & $2$ & $0.7727-0.4799i$  &$0.7727-0.4799i$ & $0.7915-0.3789i$ & $0.7727-0.4799i$ & $0.7499-0.4128i$ \\
     & $3$ & $0.7398-0.6839i$  &$0.7398-0.6839i$ & $0.7497-0.5368i$ & $0.7398-0.6839i$ & $0.6819-0.5749i$ \\
     & $4$ & $0.7015-0.8982i$   &$0.7015-0.8982i$ & $\mbox{N/A}$     & $0.7015-0.8982i$ & $0.6586-0.6815i$ \\ [0.5ex] 
 \hline
 \end{tabular}
\end{table}

\begin{table} [H]
\centering
\caption{This table details the quasinormal frequencies for gravitational perturbations (\(s = 2\)) in the context of a non-extreme Lee-Wick BH, with $q = 1$ and $q = 2$ for different settings of the parameter $\alpha$. Particularly, the third and fifth columns highlight scenarios with large \(\alpha\) values, where, as expected, the QNMs get closer to the numerical results obtained by \cite{Mamani2022EPJC} through the spectral method, specifically for a Schwarzschild BH (see third column in Table~\ref{table:7}). The fourth and sixth columns correspond to choices of the parameter $\alpha$ slightly above the corresponding value of $\alpha$ characterizing an extreme Lee-Wick BH. The numerical values were derived using the spectral method with \(300\) polynomials, which allowed to achieve an accuracy of \(300\) digits. In this context, \(\Omega\) signifies the dimensionless frequency, as introduced in equation (\ref{ourODE}). The round bracket $(q,\,\alpha)$ indicates the corresponding choice of the parameters $q$ and $\alpha$.}
\label{table:8}
 \vspace*{1em}
 \begin{tabular}{||c|c|c|c|c|c|c|c||}
 \hline
 $\ell$ & 
 $n$    & 
 $\Omega$, $(1, 100)$ & 
 $\Omega$, $(1, 2.2)$ & 
 $\Omega$, $(2, 100)$ & 
 $\Omega$, $(2, 0.7)$ &
 $\Omega$, $(2, 1.0)$ &
 $\Omega$, $(2, 2.0)$ \\ [0.5ex] 
\hline\hline
 $2$ & $0$ & $0.3737-0.0890i$ & $0.3120-0.0711i$ & $0.3737-0.0890i$  & $0.0000-0.1051i$ & $0.0000-0.2616i$ & $0.3333-0.1733i$ \\ 
     & $1$ & $0.3467-0.2739i$ & $0.2741-0.2110i$ & $0.3467-0.2739i$  & $0.0000-0.1476i$ & $0.0000-0.3050i$ & $0.0000-0.1609i$ \\
     & $2$ & $0.3011-0.4783i$ & $0.0000-0.2104i$ & $0.3011-0.4783i$  & $0.0000-0.1900i$ & $0.0000-1.1197i$ & $0.0000-3.0197i$ \\
\hline
 $3$ & $0$ & $0.5994-0.0927i$ & $0.5165-0.0836i$ & $0.5994-0.0927i$  & $0.3289-0.0807i$ & $0.3830-0.1513i$ & $0.5549-0.1780i$ \\
     & $1$ & $0.5826-0.2813i$ & $0.4770-0.2579i$ & $0.5826-0.2813i$  & $0.0000-0.1372i$ & $0.4611-0.2066i$ & $0.8575-0.3173i$ \\
     & $2$ & $0.5517-0.4791i$ & $0.4129-0.3623i$ & $0.5517-0.4791i$  & $0.0000-0.1792i$ & $0.5230-0.4444i$ & $0.0000-0.9037i$ \\
     & $3$ & $0.5112-0.6903i$ & $0.0000-0.2773i$ & $0.5112-0.6903i$  & $0.0000-0.2215i$ & $0.0000-1.2956i$ & $0.0000-3.0043i$ \\
\hline
 $4$ & $0$ & $0.8092-0.0942i$ & $0.7059-0.0887i$ & $0.8092-0.0942i$  & $0.4680-0.0829i$ & $0.6092-0.1476i$ & $0.7338-0.1798i$  \\
     & $1$ & $0.7966-0.2843i$ & $0.6791-0.2722i$ & $0.7966-0.2843i$  & $0.3365-0.1368i$ & $0.4569-0.1693i$ & $1.0119-0.2769i$  \\
     & $2$ & $0.7727-0.4799i$ & $0.5396-0.3500i$ & $0.7727-0.4799i$  & $0.0000-0.1697i$ & $0.9718-0.6008i$ & $1.1544-2.4002i$  \\
     & $3$ & $0.7398-0.6839i$ & $0.0000-0.3450i$ & $0.7398-0.6839i$  & $0.0000-0.2115i$ & $0.0000-1.6002i$ & $0.0000-7.7215i$  \\
     & $4$ & $0.7015-0.8982i$ & $0.0000-0.4183i$ & $0.7015-0.8982i$  & $0.0000-0.2535i$ & $0.0000-1.9241i$ & $0.0000-11.1020i$ \\ [0.5ex]
 \hline
 \end{tabular}
\end{table}

\begin{table} [H]
\centering
\caption{This table details the quasinormal frequencies for gravitational perturbations (\(s = 2\)) in the context of a non-extreme Lee-Wick BH, with $q = 2$ for different settings of the parameter $\alpha$. The choices of the parameter $\alpha$ are slightly above the corresponding values of $\alpha$, characterizing extreme Lee-Wick BHs of Type A and B. The numerical values were derived using the spectral method with \(300\) polynomials, which allowed to achieve an accuracy of \(300\) digits. In this context, \(\Omega\) signifies the dimensionless frequency, as introduced in equation (\ref{ourODE}). The round bracket $(q,\,\alpha)$ indicates the corresponding choice of the parameters $q$ and $\alpha$. We observe a significant deviation in the quasinormal modes compared to those of a Schwarzschild black hole, as detailed in the third column of Table~\ref{table:7}. \textbf{The occurrence of purely imaginary QNMs is accentuated for $\alpha=4.1$, i.e., close to the extremal case of Type B (see Table~\ref{tableEins}).}}
\label{table:9}
 \vspace*{1em}
 \begin{tabular}{||c|c|c|c|c|c|c||} 
 \hline
 $\ell$ & 
 $n$    & 
 $\Omega$, $(2, 3.0)$ &
 $\Omega$, $(2, 4.1)$ & 
 $\Omega$, $(2, 5.2)$ &
 $\Omega$, $(2, 6.0)$ &
 $\Omega$, $(2, 10)$ \\ [0.5ex] 
 \hline\hline
 $2$ & $0$ & $0.4162-0.1054i$ & $0.3450-0.1177i$ & $0.3856-0.1040i$ & $0.3841-0.0797i$ & $0.3732-0.0891i$ \\ 
     & $1$ & $0.0000-0.3037i$ & $0.2430-0.3123i$ & $0.0000-0.5448i$ & $0.3948-0.2811i$ & $0.3445-0.2721i$ \\
     & $2$ & $0.8625-0.3286i$ & $0.0000-0.2415i$ & $0.0000-1.1197i$ & $0.0000-0.6159i$ & $0.0000-0.9797i$ \\
\hline
 $3$ & $0$ & $0.6370-0.0905i$ & $0.5821-0.1254i$ & $0.6168-0.0989i$ & $0.6017-0.0776i$ & $0.5993-0.0931i$ \\
     & $1$ & $0.9873-0.2587i$ & $0.5294-0.7343i$ & $0.6137-0.3236i$ & $0.6227-0.2560i$ & $0.5807-0.2822i$ \\
     & $2$ & $1.2611-0.4071i$ & $0.0000-0.2734i$ & $0.6294-1.2519i$ & $0.5234-1.7905i$ & $0.5459-0.4773i$ \\ 
     & $3$ & $1.4146-0.5206i$ & $0.0000-0.4212i$ & $0.0000-2.6788i$ & $0.0000-4.6212i$ & $0.7938-1.4802i$ \\
\hline
 $4$ & $0$ & $0.8330-0.0771i$ & $0.7989-0.1301i$ & $0.8288-0.0938i$ & $0.8030-0.0783i$ & $0.8093-0.0945i$ \\
     & $1$ & $1.1622-0.2247i$ & $0.7484-0.3584i$ & $0.8404-0.3127i$ & $0.8253-0.2399i$ & $0.7959-0.2861i$ \\
     & $2$ & $1.3860-0.3747i$ & $0.9235-1.2641i$ & $1.1376-1.8826i$ & $0.8521-0.4418i$ & $0.7676-0.4824i$ \\
     & $3$ & $1.4751-0.5074i$ & $0.0000-0.3218i$ & $0.0000-2.6579i$ & $1.2851-2.4535i$ & $1.2815-1.5382i$ \\
     & $4$ & $0.0000-2.7199i$ & $0.0000-0.4630i$ & $0.0000-4.1383i$ & $0.0000-4.6813i$ & $1.1304-2.4081i$ \\ [0.5ex]
 \hline
 \end{tabular}
\end{table}

\begin{table} [H]
\centering
\caption{This table presents the quasinormal frequencies for scalar perturbations ($s = 0$) of the extreme Lee-Wick BH of type B for different choices of the parameter $q$. The results were obtained via the spectral method, employing $300$ polynomials with an accuracy of $300$ digits. Here, $\Omega$ represents the dimensionless frequency as defined in equation (\ref{ourODEext}). The numerical values of $\alpha_{e,1}$ and $\alpha_{e,2}$ are given in Table~\ref{tableEins}.}
\label{table:10}
 \vspace*{1em}
 \begin{tabular}{||c|c|c|c|c|c|c||} 
 \hline
 $\ell$ & $n$ & $\Omega$ ($q = 0.1$) & $\Omega$ ($q = 0.5$) & $\Omega$ ($q = 1.0$) & $\Omega$ ($q = 2.0, \alpha = \alpha_{e,1}$) & $\omega$ ($q = 2.0, \alpha = \alpha_{e,2}$)\\ [0.5ex] 
 \hline\hline
$0$ & $0$ & $0.1064-0.0855i$ & $0.1021-0.0881i$ & $0.0855-0.0908i$ & $\mbox{N/A}$     & $0.0918-0.1113i$ \\ 
    & $1$ & $0.0440-0.3029i$ & $0.0251-0.3102i$ & $\mbox{N/A}$     & $\mbox{N/A}$     & $\mbox{N/A}$     \\
$1$ & $0$ & $0.2971-0.0792i$ & $0.2867-0.0833i$ & $0.2495-0.0908i$ & $0.1561-0.0823i$ & $0.2765-0.1167i$ \\
    & $1$ & $0.2557-0.2443i$ & $0.2345-0.2547i$ & $\mbox{N/A}$     & $\mbox{N/A}$     & $\mbox{N/A}$     \\
    & $2$ & $0.1763-0.4328i$ & $0.1393-0.4474i$ & $\mbox{N/A}$     & $\mbox{N/A}$     & $\mbox{N/A}$     \\ 
    & $3$ & $0.0876-0.6562i$ & $\mbox{N/A}$     & $\mbox{N/A}$     & $\mbox{N/A}$     & $\mbox{N/A}$     \\
$2$ & $0$ & $0.4928-0.0787i$ & $0.4769-0.0835i$ & $0.4207-0.0936i$ & $0.2743-0.0949i$ & $0.4711-0.1231i$ \\
    & $1$ & $0.4672-0.2383i$ & $0.4431-0.2517i$ & $0.3543-0.2696i$ & $\mbox{N/A}$     & $\mbox{N/A}$     \\
    & $2$ & $0.4150-0.4050i$ & $0.3759-0.4215i$ & $\mbox{N/A}$     & $\mbox{N/A}$     & $\mbox{N/A}$     \\
    & $3$ & $0.3370-0.5836i$ & $\mbox{N/A}$ & $\mbox{N/A}$         & $\mbox{N/A}$     & $\mbox{N/A}$     \\ 
    & $4$ & $0.2495-0.7812i$ & $\mbox{N/A}$ & $\mbox{N/A}$         & $\mbox{N/A}$     & $\mbox{N/A}$     \\ [0.5ex] 
 \hline
 \end{tabular}
\end{table}

\begin{table} [H]
\centering
\caption{This table presents the quasinormal frequencies for electromagnetic perturbations ($s = 1$) of the extreme Lee-Wick BH of type B for different choices of the parameter $q$. The results were obtained via the spectral method, employing $300$ polynomials with an accuracy of $300$ digits. Here, $\Omega$ represents the dimensionless frequency as defined in equation (\ref{ourODEext}). The numerical values of $\alpha_{e,1}$ and $\alpha_{e,2}$ are given in Table~\ref{tableEins}.}
\label{table:11}
 \vspace*{1em}
  \begin{tabular}{||c|c|c|c|c|c|c||} 
 \hline
  $\ell$ & $n$ & $\Omega$ ($q = 0.1$) & $\Omega$ ($q = 0.5$) & $\Omega$ ($q = 1.0$) & $\Omega$ ($q = 2.0, \alpha = \alpha_{e,1}$) & $\omega$ ($q = 2.0, \alpha = \alpha_{e,2}$)\\ [0.5ex] 
 \hline\hline
$1$ & $0$ & $0.2588-0.0721i$ & $0.2472-0.0749i$ & $0.2092-0.0777i$ & $0.1257-0.0637i$ & $0.2266-0.1045i$ \\  
    & $1$ & $0.2177-0.2230i$ & $0.1967-0.2305i$ & $\mbox{N/A}$     & $\mbox{N/A}$     & $\mbox{N/A}$     \\ 
    & $2$ & $0.1373-0.3996i$ & $0.0982-0.4080i$ & $\mbox{N/A}$     & $\mbox{N/A}$     & $\mbox{N/A}$     \\ 
$2$ & $0$ & $0.4695-0.0759i$ & $0.4527-0.0802i$ & $0.3949-0.0877i$ & $0.2538-0.0831i$ & $0.4409-0.1197i$ \\ 
    & $1$ & $0.4444-0.2301i$ & $0.4201-0.2422i$ & $0.3336-0.2582i$ & $\mbox{N/A}$     & $\mbox{N/A}$     \\ 
    & $2$ & $0.3930-0.3919i$ & $0.3532-0.4073i$ & $\mbox{N/A}$     & $\mbox{N/A}$     & $\mbox{N/A}$     \\  
    & $3$ & $0.3150-0.5672i$ & $\mbox{N/A}$     & $\mbox{N/A}$     & $\mbox{N/A}$     & $\mbox{N/A}$     \\ 
    & $4$ & $0.2219-0.7563i$ & $\mbox{N/A}$     & $\mbox{N/A}$     & $\mbox{N/A}$     & $\mbox{N/A}$     \\ [0.5ex]
 \hline
 \end{tabular}
\end{table}

\begin{table} [H]
\centering
\caption{This table presents the quasinormal frequencies for vector-type gravitational perturbations ($s = 2$) of the extreme Lee-Wick BH of type B for different choices of the parameter $q$. The results were obtained via the spectral method, employing $300$ polynomials with an accuracy of $300$ digits. Here, $\Omega$ represents the dimensionless frequency as defined in equation (\ref{ourODEext}). The numerical values of $\alpha_{e,1}$ and $\alpha_{e,2}$ are given in Table~\ref{tableEins}.}
\label{table:12}
 \vspace*{1em}
\begin{tabular}{||c|c|c|c|c|c|c||} 
 \hline
  $\ell$ & $n$ & $\Omega$ ($q = 0.1$) & $\Omega$ ($q = 0.5$) & $\Omega$ ($q = 1.0$) & $\Omega$ ($q = 2.0, \alpha = \alpha_{e,1}$) & $\Omega$ ($q = 2.0, \alpha = \alpha_{e,2}$) \\ [0.5ex] 
 \hline\hline
$2$ & $0$ & $0.3941-0.0641i$ & $0.3743-0.0664i$ & $0.3111-0.0682i$ & $0.1724-0.0537i$ & $0.3416-0.1173i$ \\
    & $1$ & $0.3758-0.1975i$ & $0.3511-0.2044i$ & $0.2741-0.2047i$ & $\mbox{N/A}$     & $\mbox{N/A}$     \\
    & $2$ & $0.3337-0.3451i$ & $0.2952-0.3549i$ & $\mbox{N/A}$     & $\mbox{N/A}$     & $\mbox{N/A}$     \\
    & $3$ & $0.2592-0.5130i$ & $0.2159-0.5205i$ & $\mbox{N/A}$     & $\mbox{N/A}$     & $\mbox{N/A}$     \\
    & $4$ & $0.1714-0.6899i$ & $\mbox{N/A}$     & $\mbox{N/A}$     & $\mbox{N/A}$     & $\mbox{N/A}$     \\
$3$ & $0$ & $0.6200-0.0718i$ & $0.5956-0.0755i$ & $0.5143-0.0812i$ & $0.3262-0.0781i$ & $0.5790-0.1256i$ \\
    & $1$ & $0.6037-0.2172i$ & $0.5746-0.2286i$ & $0.4760-0.2510i$ & $\mbox{N/A}$     & $0.5240-0.7148i$ \\
    & $2$ & $0.5699-0.3677i$ & $0.5279-0.3865i$ & $0.4071-0.3544i$ & $\mbox{N/A}$     & $\mbox{N/A}$     \\ 
    & $3$ & $0.5159-0.5270i$ & $0.4583-0.5321i$ & $\mbox{N/A}$     & $\mbox{N/A}$     & $\mbox{N/A}$     \\
    & $4$ & $0.4346-0.6932i$ & $\mbox{N/A}$     & $\mbox{N/A}$     & $\mbox{N/A}$     & $\mbox{N/A}$     \\ 
    & $5$ & $0.3719-0.8365i$ & $\mbox{N/A}$     & $\mbox{N/A}$     & $\mbox{N/A}$     & $\mbox{N/A}$     \\ [0.5ex] 
 \hline
 \end{tabular}
\end{table}

\begin{table} [H]
\centering
\caption{This table presents the quasinormal frequencies for scalar perturbations ($s = 0$) of the extreme Lee-Wick BH of type A for $q = 2$, $\alpha_e = 5.18635$, and $x_h = 1.04013$. The results were obtained via the spectral method, employing $300$ polynomials with an accuracy of $300$ digits. Here, $\Omega$ represents the dimensionless frequency as defined in equation (\ref{ourODEext}).}
\label{table:13}
 \vspace*{1em}
 \begin{tabular}{||c|c|c||} 
 \hline
 $\ell$ & $n$ & $\Omega$ \\ [0.5ex] 
 \hline\hline
$0$ & $0$ & $0.1082-0.1159i$ \\
    & $1$ & $0.0000-1.3000i$ \\
    & $2$ & $0.0000-2.6963i$ \\
    & $3$ & $0.0000-4.0843i$ \\
    & $4$ & $0.0000-5.4686i$ \\
    & $5$ & $0.0000-6.8508i$ \\
$1$ & $0$ & $0.3011-0.1069i$ \\
    & $1$ & $0.0000-1.4239i$ \\
    & $2$ & $0.0000-2.7761i$ \\ 
    & $3$ & $0.0000-4.1443i$ \\
$2$ & $0$ & $0.4975-0.1011i$ \\
    & $1$ & $0.0000-1.6134i$ \\
    & $2$ & $0.0000-2.9125i$ \\
    & $3$ & $0.0000-4.2519i$ \\ 
    & $4$ & $0.0000-5.6066i$ \\ [0.5ex] 
 \hline
 \end{tabular}
\end{table}

\begin{table} [H]
\centering
\caption{This table presents the quasinormal frequencies for electromagnetic perturbations ($s = 1$) of the extreme Lee-Wick BH of type A for $q = 2$, $\alpha_e = 5.18635$, and $x_h = 1.04013$. The results were obtained via the spectral method, employing $300$ polynomials with an accuracy of $300$ digits. Here, $\Omega$ represents the dimensionless frequency as defined in equation (\ref{ourODEext}).}
\label{table:14}
 \vspace*{1em}
  \begin{tabular}{||c|c|c||}  
 \hline
  $\ell$ & $n$ & $\Omega$ \\ [0.5ex] 
 \hline\hline
$1$ & $0$ & $0.2513-0.1038i$ \\
    & $1$ & $0.0000-1.5146i$ \\
    & $2$ & $0.0000-2.9025i$ \\
$2$ & $0$ & $0.4702-0.1018i$ \\
    & $1$ & $0.4617-0.3327i$ \\
    & $2$ & $0.0000-1.6737i$ \\
    & $3$ & $0.0000-3.0082i$ \\
    & $4$ & $0.0000-4.3717i$ \\ [0.5ex]
 \hline
 \end{tabular}
\end{table}

\begin{table} [H]
\centering
\caption{This table presents the quasinormal frequencies for vector-type gravitational perturbations ($s = 2$) of the extreme Lee-Wick BH of type A for $q = 2$, $\alpha_e = 5.18635$, and $x_h = 1.04013$. The results were obtained via the spectral method, employing $300$ polynomials with an accuracy of $300$ digits. Here, $\Omega$ represents the dimensionless frequency as defined in equation (\ref{ourODEext}).}
\label{table:15}
 \vspace*{1em}
\begin{tabular}{||c|c|c||}
 \hline
  $\ell$ & $n$ & $\Omega$ \\ [0.5ex] 
 \hline\hline
$2$ & $0$ & $0.3854-0.1044i$ \\
    & $1$ & $0.0000-0.5451i$ \\
    & $2$ & $0.0000-1.1039i$ \\
    & $3$ & $0.0000-2.6025i$ \\
    & $4$ & $0.0000-3.9828i$ \\
$3$ & $0$ & $0.6167-0.0993i$ \\
    & $1$ & $0.6129-0.3244i$ \\
    & $2$ & $0.6293-1.2437i$ \\ 
    & $3$ & $0.0000-2.6480i$ \\
    & $4$ & $0.0000-4.0423i$ \\ 
    & $5$ & $0.0000-5.4143i$ \\ [0.5ex] 
 \hline
 \end{tabular}
\end{table}

\section{Conclusions and outlook}
\label{Sec4}

In the present work, we extensively explored the QNMs of Lee-Wick BHs for the first time, addressing scalar, electromagnetic, and gravitational perturbations through a novel application of the Spectral Method. The findings underscore the versatility of the Spectral Method in probing BH perturbations and its efficacy over previous methodologies, paving the way for further exploration and validation of theoretical predictions in the realm of BH physics and beyond. 

In our study, purely imaginary QNMs emerged in nearly extremal Lee-Wick BHs of type B, as detailed in Tables \ref{table:2}, \ref{table:3}, \ref{table:5}, \ref{table:6}, \ref{table:8}, and \ref{table:9}, becoming more prominent closer to the extremal configurations. Such modes also seem to occur for extreme Lee-Wick BHs of type A; see Tables \ref{table:13}, \ref{table:14}, and \ref{table:15}. At first sight, this result suggests that the extremality of an inner horizon could be somehow perceived outside the event horizon. Nevertheless, it could also be that the regime of extreme BHs of type A is not too far from type B, and for this reason, we still observe purely imaginary QNMs. In fact, in the tables mentioned above, we notice that these modes occur even for BH regimes in between two extremal situations.

Purely imaginary QNMs indicate an overdamped response to perturbations, suggesting a rapid return to equilibrium without oscillation. Such characteristics are similar to zero-damped modes that converge to a purely imaginary value in extremal limits, commonly observed in various BH regimes, including Reissner-Nordström, Kerr, and Kerr-Newman, among others. This phenomenon underscores the unique stability and damping behaviours of Lee-Wick BHs compared to other types, highlighting potential distinctions in horizon dynamics and stability under perturbations.
In this regard, it would be interesting to study the stability of Lee-Wick BHs under gravitational perturbations that follow higher-order differential equations in consonance with the higher-derivative nature of Lee-Wick gravity theories.

The nature of these purely imaginary QNMs might offer insights into the thermodynamic properties of Lee-Wick BHs. Given the relationship between QNMs and BH temperature (as related to surface gravity), the absence of an oscillatory component in these modes might reflect unique aspects of energy emission or absorption processes in such BHs.

It is worth mentioning that the work of \cite{Hod1998PRL, Maggiore2008PRL} linked purely imaginary QNMs to horizon quantization in the context of a Schwarzschild BH with Gauss-Bonnet corrections. In their approach, the presence of such frequencies was used to suggest discrete properties associated with the BH horizon, particularly in terms of its area and possibly its entropy. This line of reasoning is grounded in the broader context of BH quantization theories, where the quantum nature of spacetime at the BH horizon leads to a quantized spectrum. This intriguing correlation between purely imaginary QNMs and the quantum aspects of Lee-Wick BHs will be pursued further in future work. 

Last but not least, extending the idea of horizon quantization to Lee-Wick BHs could help in understanding how modifications to Einstein's gravity (as proposed by the Lee-Wick theories) affect the quantum aspects of BHs. This could provide a unique testing ground for theories that attempt to unify general relativity with quantum mechanics.

\section*{Code availability}

All analytical calculations presented in this document have been verified using the computer algebra system \textsc{Maple}. For transparency and reproducibility, we have included three \textsc{Maple} worksheets that correspond to the analyses conducted in the Sections devoted to the study of the QNMs for the non-extreme and extreme cases of Type A and B of a Lee-Wick BH. The discretization of differential operators \eqref{L0none}-\eqref{L2none} using the Chebyshev-type spectral method is equally performed in \textsc{Maple} computer algebra system. Finally, the numerical resolution of the derived quadratic eigenvalue problem, denoted by equation \eqref{eq:eig}, is executed in the \textsc{Matlab} environment utilizing the \texttt{polyeig} function. Access to all these resources is provided through the following repository link, ensuring that interested parties can freely review and utilize the computational methodologies employed in our study:

\begin{itemize}
    \item \url{https://github.com/dutykh/LeeWick/}
\end{itemize}

\section*{Acknowledgements}

This publication is based upon work supported by the Khalifa University of Science and Technology under Award No. FSU$-2023-014$. B.L.G. is supported by Primus grant PRIMUS/23/SCI/005 from Charles University.

\appendix
\section{Expressions for the limits appearing in Table~\ref{tableNONEXT}}\label{AppendixA}
\begin{equation}
\Lambda_1=\frac{a_{1,1}\cos{(2\alpha x_h q)}+b_{1,1}\sin{(2\alpha x_h q)}+c_{1,1}}{\widetilde{a}_{1,1}\cos{(2\alpha x_h q)}+\widetilde{b}_{1,1}\sin{(2\alpha x_h q)}+\widetilde{c}_{1,1}}
\end{equation}
with
\begin{eqnarray}
a_{1,1}&=&\frac{x_h}{2}(\alpha^3 q^6 x_h^3-\alpha^3 q^6 x_h^2+3\alpha^3 q^4 x_h^3-2\alpha^2 q^6 x_h^2-3\alpha^3 q^4 x_h^2+\alpha^2 q^6 x_h+3\alpha^3 q^2 x_h^3-2\alpha^2 q^4 x_h^2-3\alpha^3 q^2 x_h^2\nonumber\\
&&+\alpha^2 q^4 x_h+\alpha^3 x_h^3+2\alpha^2 q^2 x_h^2-6\alpha q^4 x_h-\alpha^3 x_h^2-\alpha^2 q^2 x_h+2\alpha^2 x_h^2-4\alpha q^2 x_h+q^4-\alpha^2 x_h\nonumber\\
&&+2\alpha x_h-6q^2+1),\\
b_{1,1}&=&\frac{x_h}{2}(-\alpha^3 q^7 x_h^3+\alpha^3 q^7 x_h^2-3\alpha^3 q^5 x_h^3+3\alpha^3 q^5 x_h^2-3\alpha^3 q^3 x_h^3-4\alpha^2 q^5 x_h^2+3\alpha^3 q^3 x_h^2+2\alpha^2 q^5 x_h-\alpha^3 q x_h^3\nonumber\\
&&-8\alpha^2 q^3 x_h^2+2\alpha q^5 x_h+\alpha^3 q x_h^2+4\alpha^2 q^3 x_h-4\alpha^2 q x_h^2-4\alpha q^3 x_h+2\alpha^2 q x_h-6\alpha q x_h+4q^3-4q),\\
c_{1,1} &=&\frac{x_h}{2}(-\alpha^3 q^6 x_h^3+\alpha^3 q^6 x_h^2 3\alpha^3 q^4 x_h^3+3\alpha^3 q^4 x_h^2-\alpha^2 q^6 x_h-3\alpha^3 q^2 x_h^3-2\alpha^2 q^4 x_h^2+3\alpha^3 q^2 x_h^2-\alpha^2 q^4 x_h\nonumber\\
&&-\alpha^3 x_h^3-4\alpha^2 q^2 x_h^2-2\alpha q^4 x_h+\alpha^3 x_h^2+\alpha^2 q^2 x_h-2\alpha^2 x_h^2-4\alpha q^2 x_h-q^4+\alpha^2 x_h-2\alpha x_h-2q^2-1),\\
\widetilde{a}_{1,1}&=&\frac{1}{2}(\alpha q^3 x_h+\alpha q^2 x_h+\alpha q x_h+\alpha x_h-q^2+2q+1)(\alpha q^3 x_h-\alpha q^2 x_h+\alpha q x_h-\alpha x_h+q^2+2q-1),\\
\widetilde{b}_{1,1}&=&q(\alpha q^2 x_h+\alpha x_h-q^2+1)(\alpha q^2 x_h+\alpha x_h+2),\\ 
\widetilde{c}_{1,1}&=&\frac{1}{2}(q^2+1)^2 (\alpha^2 q^2 x_h^2+\alpha^2 x_h^2+2\alpha x_h+1).
\end{eqnarray}
\begin{equation}
    \Lambda_0=\frac{a_{1,0}\cos{(2\alpha x_h q)}+b_{1,0}\sin{(2\alpha x_h q)}+c_{1,0}}{\widetilde{a}_{1,0}\cos{(2\alpha x_h q)}+\widetilde{b}_{1,0}\sin{(2\alpha x_h q)}+\widetilde{c}_{1,0}}
\end{equation}
with
\begin{eqnarray}
a_{1,0}&=&-\frac{1}{2}\left(\alpha^2 q^4 x_h^2 \alpha^2 q^4 x_h+2\alpha^2 q^2 x_h^2-2\alpha^2 q^2 x_h-\alpha q^3 x_h+\alpha^2 x_h^2+\alpha q^2 x_h-\alpha^2 x_h-\alpha q x_h+\alpha x_h-q^2-2q+1\right)\cdot\nonumber\\
&&\left(\alpha^2 q^4 x_h^2-\alpha^2 q^4 x_h+2\alpha^2 q^2 x_h^2 2\alpha^2 q^2 x_h+\alpha q^3 x_h+\alpha^2 x_h^2+\alpha q^2 x_h-\alpha^2 x_h+\alpha q x_h+\alpha x_h-q^2+2q+1\right),\\
b_{1,0}&=&q(\alpha q^2 x_h+\alpha x_h+2)(\alpha^2 q^4 x_h^2-\alpha^2 q^4 x_h+2\alpha^2 q^2 x_h^2-2\alpha^2 q^2 x_h+\alpha^2 x_h^2+\alpha q^2 x_h-\alpha^2 x_h+\alpha x_h-q^2+1),\\
c_{1,0}&=&\frac{1}{2}(q^2+1)^2\left(\alpha^4 q^4 x_h^4-2\alpha^4 q^4 x_h^3+\alpha^4 q^4 x_h^2+2\alpha^4 q^2 x_h^4-4\alpha^4 q^2 x_h^3+2\alpha^4 q^2 x_h^2+\alpha^4 x_h^4+2\alpha^3 q^2 x_h^3-2\alpha^4 x_h^3\right.\nonumber\\
&&\left.-2\alpha^3 q^2 x_h^2+\alpha^4 x_h^2+2\alpha^3 x_h^3-\alpha^2 q^2 x_h^2-2\alpha^3 x_h^2+2\alpha^2 q^2 x_h+3\alpha^2 x_h^2-2\alpha^2 x_h+2\alpha x_h+1\right),\\
\widetilde{a}_{1,0}&=&4\widetilde{a}_{1,1},\quad
\widetilde{b}_{1,0}=4\widetilde{b}_{1,1},\quad
\widetilde{c}_{1,0}=4\widetilde{c}_{1,1}.
\end{eqnarray}
\begin{equation}
    A_2=\frac{d_1\cos{(\alpha x_h q)}+e_1\sin{(\alpha x_h q)}}{\widetilde{d}_1\cos{(\alpha x_h q)}+\widetilde{e}_1\sin{(\alpha x_h q)}}
\end{equation}
with
\begin{eqnarray}
d_1&=&q(1+x_h^2)(\alpha q^2 x_h+\alpha x_h+2),\\ 
e_1&=&\alpha^2 q^4 x_h^2-\alpha^2 q^4 x_h+2\alpha^2 q^2 x_h^2+\alpha q^2 x_h^3-2\alpha^2 q^2 x_h+\alpha^2 x_h^2+\alpha q^2 x_h\nonumber\\
&&+\alpha x_h^3-q^2 x_h^2-\alpha^2 x_h+\alpha x_h-q^2+x_h^2+1,\\
\widetilde{d}_1&=&q(\alpha q^2 x_h+\alpha x_h+2),\\
\widetilde{e}_1&=&\alpha^2 q^4 x_h^2-\alpha^2 q^4 x_h+2\alpha^2 q^2 x_h^2-2\alpha^2 q^2 x_h+\alpha^2 x_h^2+\alpha q^2 x_h-\alpha^2 x_h+\alpha x_h-q^2+1.
\end{eqnarray}
\begin{equation}
    A_0=-\frac{\ell}{8}(\ell+1).
\end{equation}
Moreover, we have
\begin{equation}
B_2=\frac{d_{2,2}\cos{(2\alpha x_h q)}+e_{2,2}\sin{(2\alpha x_h q)}+f_{2,2}}{\widetilde{d}_{2,2}\cos{(2\alpha x_h q)}+\widetilde{e}_{2,2}\sin{(2\alpha x_h q)}+\widetilde{f}_{2,2}}
\end{equation}
with
\begin{eqnarray}
d_{2,2}&=&\frac{x_h}{2}\left(3\alpha^4 q^8 x_h^5-3\alpha^4 q^8 x_h^4-2\alpha^4 q^8 x_h^3+12\alpha^4 q^6 x_h^5+2\alpha^4 q^8 x_h^2-12\alpha^4 q^6 x_h^4-8\alpha^4 q^6 x_h^3+18\alpha^4 q^4 x_h^5+8\alpha^3 q^6 x_h^4\right.\nonumber\\
&&\left.+8\alpha^4 q^6 x_h^2-18\alpha^4 q^4 x_h^4-4\alpha^3 q^6 x_h^3-12\alpha^4 q^4 x_h^3+12\alpha^4 q^2 x_h^5-4\alpha^3 q^6 x_h^2+24\alpha^3 q^4 x_h^4-13\alpha^2 q^6 x_h^3+12\alpha^4 q^4 x_h^2\right.\nonumber\\
&&\left.-12\alpha^4 q^2 x_h^4-12\alpha^3 q^4 x_h^3+\alpha^2 q^6 x_h^2-8\alpha^4 q^2 x_h^3+3\alpha^4 x_h^5-12\alpha^3 q^4 x_h^2+24\alpha^3 q^2 x_h^4+4\alpha^2 q^6 x_h-13\alpha^2 q^4 x_h^3\right.\nonumber\\
&&\left.+8\alpha^4 q^2 x_h^2-3\alpha^4 x_h^4-12\alpha^3 q^2 x_h^3+\alpha^2 q^4 x_h^2-2\alpha^4 x_h^3-12\alpha^3 q^2 x_h^2+8\alpha^3 x_h^4+4\alpha^2 q^4 x_h+13\alpha^2 q^2 x_h^3-36\alpha q^4 x_h^2\right.\nonumber\\
&&\left.+2\alpha^4 x_h^2-4\alpha^3 x_h^3-\alpha^2 q^2 x_h^2-12\alpha q^4 x_h-4\alpha^3 x_h^2-4\alpha^2 q^2 x_h+13\alpha^2 x_h^3-24\alpha q^2 x_h^2+6 q^4 x_h-\alpha^2 xh^2\right.\nonumber\\
&&\left.-8\alpha q^2 x_h+2q^4-4\alpha^2 x_h+12\alpha x_h^2-36q^2 x_h+4\alpha x_h-12q^2+6x_h+2\right),\\
e_{2,2}&=&-qx_h\left(4\alpha^3 q^6 x_h^4-2\alpha^3 q^6 x_h^3-2\alpha^3 q^6 x_h^2+12\alpha^3 q^4 x_h^4-6\alpha^3 q^4 x_h^3-6\alpha^3 q^4 x_h^2+12\alpha^3 q^2 x_h^4+13\alpha^2 q^4 x_h^3-6\alpha^3 q^2 x_h^3\right.\nonumber\\
&&\left.-\alpha^2 q^4 x_h^2-6\alpha^3 q^2 x_h^2+4\alpha^3 x_h^4-4\alpha^2 q^4 x_h+26\alpha^2 q^2 x_h^3-6\alpha q^4 x_h^2-2\alpha^3 x_h^3-2\alpha^2 q^2 x_h^2-2\alpha q^4 x_h-2\alpha^3 x_h^2-8\alpha^2 q^2 x_h\right.\nonumber\\
&&\left.+13\alpha^2 x_h^3+12\alpha q^2 x_h^2-\alpha^2 x_h^2+4\alpha q^2 x_h-4\alpha^2 xh+18\alpha x_h^2-12q^2 x_h+6\alpha x_h-4q^2+12xh+4\right),\\
f_{2,2}&=&-\frac{x_h}{2}(q^2+1)^2\left(\alpha^4 q^4 x_h^5-\alpha^4 q^4 x_h^4-2\alpha^4 q^4 x_h^3+4\alpha^4 q^2 x_h^5+2\alpha^4 q^4 x_h^2-4\alpha^4 q^2 x_h^4-4\alpha^4 q^2 x_h^3+3\alpha^4 x_h^5+4\alpha^3 q^2 x_h^4\right.\nonumber\\
&&\left.+4\alpha^4 q^2 x_h^2-3\alpha^4 x_h^4-2\alpha^4 x_h^3-4\alpha^3 q^2 x_h^2+8\alpha^3 x_h^4-\alpha^2 q^2 x_h^3+2\alpha^4 x_h^2-4\alpha^3 x_h^3+5\alpha^2 q^2 x_h^2-4\alpha^3 x_h^2\right.\nonumber\\
&&\left.+4\alpha^2 q^2 x_h+13\alpha^2 x_h^3-\alpha^2 x_h^2-4\alpha^2 x_h+12\alpha x_h^2+4\alpha x_h+6 x_h+2\right),\\
\widetilde{d}_{2,2}&=&\alpha^3 q^6 x_h^3-\alpha^3 q^6 x_h^2+3\alpha^3 q^4 x_h^3-2\alpha^2 q^6 x_h^2-3\alpha^3 q^4 x_h^2+\alpha^2 q^6 x_h+3\alpha^3 q^2 x_h^3-2\alpha^2 q^4 x_h^2-3\alpha^3 q^2 x_h^2+\alpha^2 q^4 x_h\nonumber\\
&&+\alpha^3 x_h^3+2\alpha^2 q^2 x_h^2-6\alpha q^4 x_h-\alpha^3 x_h^2-\alpha^2 q^2 x_h+2\alpha^2 x_h^2-4\alpha q^2 x_h+q^4-\alpha^2 x_h+2\alpha x_h-6 q^2+1,\\
\widetilde{e}_{2,2}&=&-q\left(\alpha q^2 x_h+\alpha x_h+2\right)\cdot\nonumber\\
&&\left(\alpha^2 q^4 x_h^2-\alpha^2 q^4 x_h+2\alpha^2 q^2 x_h^2-2\alpha^2 q^2 x_h+\alpha^2 x_h^2+2\alpha q^2 x_h-\alpha^2 xh+2\alpha x_h-2 q^2+2\right),\\
\widetilde{f}_{2,2}&=&-(q^2+1)^2\left(\alpha^3 q^2 x_h^3-\alpha^3 q^2 x_h^2+\alpha^3 x_h^3-\alpha^3 x_h^2+\alpha^2 q^2 x_h+2\alpha^2 x_h^2-\alpha^2 x_h+2\alpha x_h+1\right).
\end{eqnarray}

\begin{equation}
B_1=\frac{d_{2,1}\cos{(2\alpha x_h q)}+e_{2,1}\sin{(2\alpha x_h q)}+f_{2,1}}{\widetilde{d}_{2,1}\cos{(2\alpha x_h q)}+\widetilde{e}_{2,1}\sin{(2\alpha x_h q)}+\widetilde{f}_{2,1}}
\end{equation}
with
\begin{eqnarray}
d_{2,1}&=&-1+12\alpha q^2 x_h^2+2\alpha^3 q^6 x_h^3-\frac{13}{2}\alpha^2 q^2 x_h^3+2\alpha^3 q^6 x_h^2+6\alpha^3 q^2 x_h^3+\frac{1}{2}\alpha^2 q^2 x_h^2-12\alpha^3 q^4 x_h^4+6\alpha q^4 x_h -12\alpha^3 q^2 x_h^4\nonumber\\
&&+6\alpha^3 q^4 x_h^2+6\alpha^3 q^2 x_h^2-4\alpha^4 q^6 x_h^2-\alpha^4 q^8 x_h^2+\alpha^4 q^8 x_h^3-\frac{1}{2}\alpha^2 q^6 x_h^2+\frac{3}{2}\alpha^4 q^8 x_h^4+4\alpha^4 q^6 x_h^3-2\alpha^2 q^6 x_h+6\alpha^4 q^6 x_h^4\nonumber\\
&&-\frac{3}{2}\alpha^4 x_h^5+\frac{3}{2}\alpha^4 x_h^4+\alpha^4 x_h^3-\alpha^4 x_h^2+\frac{1}{2}\alpha^2 x_h^2-4\alpha^3 x_h^4+2\alpha^3 x_h^3+2\alpha^3 x_h^2-\frac{13}{2}\alpha^2 x_h^3+2\alpha^2 x_h-6\alpha x_h^2+18 q^2 x_h\nonumber\\
&&-2\alpha x_h-\frac{3}{2}\alpha^4 q^8 x_h^5-6\alpha^4 q^6 x_h^5+\frac{13}{2}\alpha^2 q^6 x_h^3-3q^4 x_h-3x_h+6q^2-q^4+\frac{13}{2}\alpha^2 q^4 x_h^3+2\alpha^2 q^2 x_h-2\alpha^2 q^4 x_h-4\alpha^3 q^6 x_h^4\nonumber\\
&&+6\alpha^3 q^4 x_h^3-6\alpha^4 q^2 x_h^5-\frac{1}{2}\alpha^2 q^4 x_h^2+4\alpha q^2 x_h+18\alpha q^4 x_h^2+4\alpha^4 q^2 x_h^3-9\alpha^4 q^4 x_h^5+6\alpha^4 q^2 x_h^4+6\alpha^4 q^4 x_h^3-6\alpha^4 q^4 x_h^2\nonumber\\
&&-4\alpha^4 q^2 x_h^2+9\alpha^4 q^4 x_h^4,\\
e_{2,1}&=&q\left(4\alpha^3 q^6 x_h^4-2\alpha^3 q^6 x_h^3-2\alpha^3 q^6 x_h^2+12\alpha^3 q^4 x_h^4-6\alpha^3 q^4 x_h^3-6\alpha^3 q^4 x_h^2+12\alpha^3 q^2 x_h^4+13\alpha^2 q^4 x_h^3-6\alpha^3 q^2 x_h^3\right.\nonumber\\
&&\left.-\alpha^2 q^4 x_h^2-6\alpha^3 q^2 x_h^2+4\alpha^3 x_h^4-4\alpha^2 q^4 x_h+26\alpha^2 q^2 x_h^3-6\alpha q^4 x_h^2-2\alpha^3 x_h^3-2\alpha^2 q^2 x_h^2-2\alpha q^4 x_h-2\alpha^3 x_h^2\right.\nonumber\\
&&\left.-8\alpha^2 q^2 xh+13\alpha^2 x_h^3+12\alpha q^2 x_h^2-\alpha^2 x_h^2+4\alpha q^2 x_h-4\alpha^2 x_h+18\alpha x_h^2-12 q^2 x_h+6\alpha x_h-4q^2+12x_h+4\right),\\
f_{2,1}&=&\frac{1}{2}(q^2+1)^2\left(\alpha^4 q^4 x_h^5-\alpha^4 q^4 x_h^4-2\alpha^4 q^4 x_h^3+4\alpha^4 q^2 x_h^5+2\alpha^4 q^4 x_h^2-4\alpha^4 q^2 x_h^4-4\alpha^4 q^2 x_h^3+3\alpha^4 x_h^5+4\alpha^3 q^2 x_h^4\right.\nonumber\\
&&\left.+4\alpha^4 q^2 x_h^2-3\alpha^4 x_h^4-2\alpha^4 x_h^3-4\alpha^3 q^2 x_h^2+8\alpha^3 x_h^4-\alpha^2 q^2 x_h^3+2\alpha^4 x_h^2-4\alpha^3 x_h^3+5\alpha^2 q^2 x_h^2-4\alpha^3 x_h^2\right.\nonumber\\
&&\left.+4\alpha^2 q^2 x_h+13\alpha^2 x_h^3-\alpha^2 x_h^2-4\alpha^2 x_h+12\alpha x_h^2+4\alpha x_h+6x_h+2\right),\\
\widetilde{d}_{2,1}&=&8\widetilde{a}_{1,1},\quad
\widetilde{e}_{2,1}=8\widetilde{b}_{1,1},\quad
\widetilde{f}_{2,1}=8\widetilde{c}_{1,1}.
\end{eqnarray}

\begin{equation}
B_0=\frac{d_{2,0}\cos{(2\alpha x_h q)}+e_{2,0}\sin{(2\alpha x_h q)}+f_{2,0}}{\widetilde{d}_{2,0}\cos{(2\alpha x_h q)}+\widetilde{e}_{2,0}\sin{(2\alpha x_h q)}+\widetilde{f}_{2,0}}
\end{equation}
with
\begin{eqnarray}
d_{2,0}&=&\frac{\epsilon}{2}\left(\alpha^2 q^4 x_h^2-\alpha^2 q^4 x_h+2\alpha^2 q^2 x_h^2-2\alpha^2 q^2 x_h-\alpha q^3 x_h+\alpha^2 x_h^2+\alpha q^2 x_h-\alpha^2 x_h-\alpha q x_h+\alpha x_h-q^2-2q+1\right)\cdot\nonumber\\
&&\left(\alpha^2 q^4 x_h^2-\alpha^2 q^4 x_h+2\alpha^2 q^2 x_h^2-2\alpha^2 q^2 x_h+\alpha q^3 x_h+\alpha^2 xh^2+\alpha q^2 x_h-\alpha^2 x_h+\alpha q x_h+\alpha x_h-q^2+2q+1\right)\nonumber\\
&&+\frac{1}{2}\ell(\ell+1)\left(\alpha^3 q^6 x_h^3-\alpha^3 q^6 x_h^2+3\alpha^3 q^4 x_h^3-2\alpha^2 q^6 x_h^2-3\alpha^3 q^4 x_h^2+\alpha^2 q^6 x_h+3\alpha^3 q^2 x_h^3-2\alpha^2 q^4 x_h^2-3\alpha^3 q^2 x_h^2\right.\nonumber\\
&&\left.+\alpha^2 q^4 x_h+\alpha^3 x_h^3+2\alpha^2 q^2 x_h^2-6\alpha q^4 x_h-\alpha^3 x_h^2-\alpha^2 q^2 x_h+2\alpha^2 x_h^2-4\alpha q^2 x_h+q^4-\alpha^2 x_h\right.\nonumber\\
&&\left.+2\alpha x_h-6q^2+1\right),\\
e_{2,0}&=&-\epsilon q\left(\alpha q^2 x_h+\alpha x_h+2\right)\left(\alpha^2 q^4 x_h^2-\alpha^2 q^4 x_h+2\alpha^2 q^2 x_h^2-2\alpha^2 q^2 x_h+\alpha^2 x_h^2+\alpha q^2 x_h-\alpha^2 x_h+\alpha x_h-q^2+1\right)\nonumber\\
&&-\frac{1}{2}q\ell(\ell+1)\left(\alpha q^2 x_h+\alpha x_h+2\right)\left(\alpha^2 q^4 x_h^2-\alpha^2 q^4 x_h+2\alpha^2 q^2 x_h^2-2\alpha^2 q^2 x_h+\alpha^2 x_h^2+2\alpha q^2 x_h-\alpha^2 x_h\right.\nonumber\\
&&\left.+2\alpha x_h-2q^2+2\right),\\
f_{2,0}&=&-\frac{\epsilon}{2}(q^2+1)^2\left(\alpha^4 q^4 x_h^4-2\alpha^4 q^4 x_h^3+\alpha^4 q^4 x_h^2+2\alpha^4 q^2 x_h^4-4\alpha^4 q^2 x_h^3+2\alpha^4 q^2 x_h^2+\alpha^4 x_h^4+2\alpha^3 q^2 x_h^3-2\alpha^4 x_h^3\right.\nonumber\\
&&\left.-2\alpha^3 q^2 x_h^2+\alpha^4 x_h^2+2\alpha^3 x_h^3-
\alpha^2 q^2 x_h^2-2\alpha^3 x_h^2+2\alpha^2 q^2 x_h+3\alpha^2 x_h^2-2\alpha^2 x_h+2\alpha x_h+1\right)-\frac{1}{2}(q^2+1)^2\ell(\ell+1)\cdot\nonumber\\
&&\left(\alpha^3 q^2 x_h^3-\alpha^3 q^2 x_h^2+\alpha^3 x_h^3-\alpha^3 x_h^2+\alpha^2 q^2 x_h+2\alpha^2 x_h^2-\alpha^2 x_h+2\alpha x_h+1\right),\\
\widetilde{d}_{2,0}&=&8\widetilde{a}_{1,1},\quad
\widetilde{e}_{2,0}=8\widetilde{b}_{1,1},\quad
\widetilde{f}_{2,0}=8\widetilde{c}_{1,1}.
\end{eqnarray}

\section{Expressions for the limits appearing in Table~\ref{tableEXT}}
\label{AppendixB}

\begin{eqnarray}
\Lambda_{1e}&=&-\frac{x_e\left[x_e^2(x_e-1)(q^2+1)^2\alpha_e^3+2x_e^2(q^2+1)\alpha_e^2-2x_e(q^2-1)\alpha_e-2\right]}{4\left[x_e(q^2+1)\alpha_e+2\right]},\\
A_{0e}&=&-\frac{\ell(\ell+1)}{16},\quad
B_{0e}=\frac{\ell(\ell+1)\Lambda_{1e}}{8x_e}.
\end{eqnarray}
\begin{equation}
A_{2e}=\frac{\sum_{n=0}^6 c_n\alpha_e^n}{6\left[x_e^2(x_e-1)(q^2+1)^2\alpha_e^3+2x_e^2(q^2+1)\alpha_e^2-2x_e(q^2-1)\alpha_e-2\right]^2}   
\end{equation}
with
\begin{eqnarray}
c_6&=&3x_e^4(x_e-1)^2(q^2+1)^4,\quad
c_5=4x_e^4(x_e-1)(x_e^2+3)(q^2+1)^3,\\
c_4&=&2x_e^3(q^2+1)^2\left[(3q^2+11)x_e^3-4(q^2+2)x_e^2+3(4-2q^2)x_e+6(q^2-1)\right],\\
c_3&=&4x_e^2(q^2+1)\left[(5q^2+13)x_e^3-4(q^2+1)x_e^2+3(1-3q^2)x_e+3(q^2+1)\right],\\
c_2&=&-4x_e^2\left[(5q^4-2q^2-15)x_e^2+3(1+4q^2-q^4)\right],\quad
c_1=-8x_e\left[(8x_e^2-3)q^2+3\right],\quad
c_0=12(1-4x_e^2).
\end{eqnarray}


\end{document}